\documentclass[titlepage,oneside,11pt]{article}

\usepackage[tiny,rm,pagestyles]{titlesec}
\titleformat{\section}{\bfseries}{\Large \thesection. }{0pt}{\Large}
\titlespacing*{\section}{0pt}{12pt}{*0}
\titleformat{\subsection}{\bfseries}{\large \thesubsection. }{0pt}{\large}
\titlespacing*{\subsection}{0pt}{12pt}{*0}
\titleformat{\subsubsection}{\bfseries}{\large \thesubsubsection. }{0pt}{\large}
\titlespacing*{\subsubsection}{0pt}{12pt}{*0}

\usepackage[utf8]{inputenc}
\usepackage[T1]{fontenc}
\usepackage{textcomp}
\usepackage{mathptmx} 
\usepackage{bm}

\usepackage[left=3cm, right=3cm, top=2.5cm, bottom=2.5cm]{geometry}
\usepackage{setspace}
\usepackage{ragged2e}
\justifying

\usepackage{fancyhdr}
\usepackage{graphicx}
\usepackage{caption}
\usepackage{subcaption}
\usepackage{float}
\usepackage{lscape}
\usepackage{rotating}
\usepackage{longtable}
\usepackage{array}
\usepackage{booktabs}
\usepackage{multirow}
\usepackage{bigstrut}
\usepackage{colortbl}
\usepackage{tabularx}
\usepackage{dcolumn}
\newcolumntype{d}[1]{D{.}{.}{#1}}
\newcolumntype{Y}{>{\centering\arraybackslash}X}

\makeatletter
\renewcommand{\fnum@figure}{\textbf{FIGURE~\thefigure}}
\renewcommand{\fnum@table}{\textbf{TABLE~\thetable}}
\makeatother

\usepackage{amsmath}
\usepackage{dsfont}

\usepackage{xcolor,soul}
\sethlcolor{yellow}

\usepackage[mathlines,pagewise]{lineno}

\usepackage{natbib}
\usepackage[bookmarksopen,bookmarksnumbered,colorlinks,linkcolor=blue,citecolor=blue]{hyperref}
\usepackage{orcidlink}

\usepackage{appendix}

\usepackage{parskip}

\makeatletter
\renewcommand\section{\@startsection {section}{1}{1em}%
                                   {-3.5ex \@plus -1ex \@minus -.2ex}%
                                   {2.3ex \@plus.2ex}%
                                   {\normalfont\fontsize{12}{14}\bfseries}}
\renewcommand{\thesection}{\arabic{section}.}
\renewcommand\subsection{\@startsection {subsection}{2}{1em}%
                                   {-3.25ex \@plus -1ex \@minus -.2ex}%
                                   {1.5ex \@plus .2ex}%
                                   {\normalfont\fontsize{12}{14}\bfseries}}
\renewcommand{\thesubsection}{\thesection\arabic{subsection}}
\makeatother

\begin{document}

	\begin{titlepage}
		
		\begin{center}

		    \LARGE{\bfseries{Beyond utility: incorporating eye-tracking, skin conductance and heart rate data into cognitive and econometric travel behaviour models}}
		    
		\end{center}
		
		\vspace{1cm}

		\begin{flushleft}
			
			{\bfseries Thomas O. Hancock (Corresponding Author)} \\
			Choice Modelling Centre \& Institute for Transport Studies \\
            34-40 University Road \\
			University of Leeds \\
            LS2 9JT \\
			T.O.Hancock@leeds.ac.uk\\[12pt]

			{\bfseries Stephane Hess}\\
			Choice Modelling Centre \& Institute for Transport Studies \\
			University of Leeds \\
			S.Hess@leeds.ac.uk\\[12pt]
		
		    {\bfseries Charisma F. Choudhury}\\
			Choice Modelling Centre \& Institute for Transport Studies \\
			University of Leeds \\
			C.F.Choudhury@leeds.ac.uk\\[12pt]
            \end{flushleft}
    \end{titlepage}			
			

        \Large{\bfseries{Abstract}}\\
			\normalsize
            \justifying
		\noindent Choice models for large-scale applications have historically relied on economic theories (e.g. utility maximisation) that establish relationships between the choices of individuals, their characteristics, and the attributes of the alternatives. In a parallel stream, choice models in cognitive psychology have focused on modelling the decision-making process, but typically in controlled scenarios. Recent research developments have attempted to bridge the modelling paradigms, with choice models that are based on psychological foundations, such as decision field theory (DFT), outperforming traditional econometric choice models for travel mode and route choice behaviour. The use of physiological data, which can provide indications about the choice-making process and mental states, opens up the opportunity to further advance the models. In particular, the use of such data to enrich 'process' parameters within a cognitive theory-driven choice model has not yet been explored. This research gap is addressed by incorporating physiological data into both econometric and DFT models for understanding decision-making in two different contexts: stated-preference responses (static) of accomodation choice and gap-acceptance decisions within a driving simulator experiment (dynamic). Results from models for the static scenarios demonstrate that both models can improve substantially through the incorporation of eye-tracking information. Results from models for the dynamic scenarios suggest that stress measurement and eye-tracking data can be linked with process parameters in DFT, resulting in larger improvements in comparison to simpler methods for incorporating this data in either DFT or econometric models. The findings provide insights into the value added by physiological data as well as the performance of different candidate modelling frameworks for integrating such data.
			

	\normalsize
	\begin{flushleft}

    	\section{Introduction}
        \justifying
        Travel behaviour modellers have typically used survey data and standard econometric choice models such as logit \citep{mcfadden1974conditional} as their main model framework for a wide range of activity-travel applications. These range from modelling long-term decisions like residential location, medium-term decisions like vehicle ownership, short-term decisions like destination, route and mode choice to very quick decisions like gap-acceptance during driving. The survey data is used to gather information about the choices, the attributes of the alternatives and characteristics of the decision-makers (income, gender and attitudes, for example). It is, however, challenging to collect information about the decision-making process or protocol via surveys. In recent research, some auxiliary data, such as choice response time, have been used as indicators of the decision-making process \citep{hess2013linking,uggeldahl2016choice}. Furthermore, literature in psychology provides evidence that choices are also affected by the `state of mind' - stress, time pressure, anger, etc. (e.g. \citealt{svenson1993time,maule2000effects,litvak2010fuel}). For example, a stressed traveller may choose to use public transport instead of driving; an angry driver may drive aggressively and decide to make more risky overtaking manoeuvres, etc. These effects have been captured in driving simulator and virtual reality studies, where e.g. stressed drivers accept smaller gaps \citep{paschalidis2018modelling} or choose alternative modes for subsequent trips \citep{henriquez4988311modelling}.
        
        It is also evident that the decision-making process (for example, mental deliberation) has a substantial impact on the final choices \citep{peters2007adult,gigerenzer2011heuristic,huang2015age}. For example, travellers may choose differently when they need to make a quick decision \citep{rendon2016effects}. Alternatively, older drivers may take longer to process information and react \citep{salvia2016effects}. Recent developments have also allowed for the incorporation of `instant utility' (momentary emotions perceived during an experience) into choice models \citep{henriquez2025identifying,henriquez2025experience}. 
        
        These notions, coupled with technological advances that enable the collection of more detailed data, have motivated behavioural modellers to gather further insights into the choice process itself through the collection of physiological data, including for trips made in real-world settings \citep{barria2023relating}. This includes but is not limited to more sophisticated data sources such as heart-rate, eye-tracking, facial expression, electroencephalogram (EEG) data and functional magnetic resonance imaging (fMRI) data. Proper utilisation of these data sources, however, warrants rethinking the framework of traditional choice models \cite{hancock2023utilising}. A typical econometric choice model may take inputs such as the attributes of alternatives, socio-demographic characteristics, attitudes and perceptions. Then, with the use of estimated utility parameters, returns probabilities for observing the outputs (choices). This is depicted in the upper panel of Figure \ref{fig:StandardModels}. The known values (represented by rectangles) are the inputs and the observed choices. Latent variables (represented by ovals) such as attitudes and perceptions can also be used if the modeller uses an Integrated Choice and Latent Variable (ICLV) model \citep{ben2002hybrid,vij2016and}. Existing econometric choice models can be used to determine the probabilities of observed choices being made given a set of alternatives and some estimated utility parameters (estimated values are represented by shaded rectangles in Figure \ref{fig:StandardModels}).
        
        \begin{figure}[ht!]
        	\centering
        	\caption{Standard econometric and psychological choice modelling frameworks.}
        	\label{fig:StandardModels}	\includegraphics[scale=0.475]{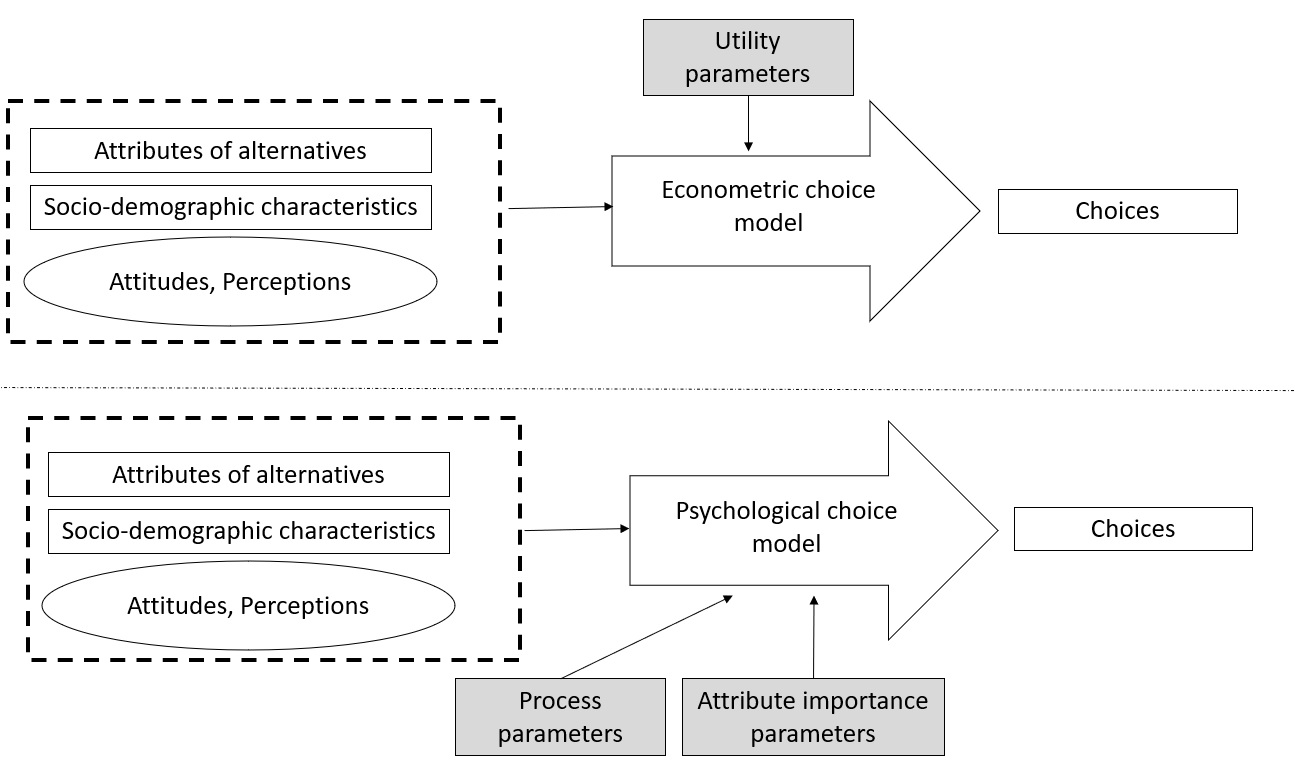}
        \end{figure}
        
        Thus far, there has been limited guidance on how data depicting the choice deliberation process (EEG, fMRI, eye-tracking, etc.) should be incorporated into these models, as these types of data do not fit intuitively into the categories of input or output variables. Whilst it has been widely established that the use of this data as explanatory variables can lead to biases in model outputs \citep{bansal2024discrete}, the use of this information within an ICLV framework \citep{krucien2017visual,bogacz2019modelling,paschalidis2019combining}, is not straightforward, partially as a result of a lack of established frameworks for how latent variables should impact choices in such settings. As a contrast, models developed by mathematical psychologists (introduced in detail in the subsequent section) are often specifically designed to capture or represent the choice deliberation process. This means that these models crucially additionally have distinct \textit{process parameters} that are included either to create a plausible psychological or neurological representation of decision-making or to allow the model to predict contextual effects that have been found by research in cognitive psychology and/or behavioural economics \citep{busemeyer2014psychological}. Previous work has demonstrated that these models can also include parameters for the relative importance of different attributes, thus allowing them to compete with standard econometric choice models \citet{hancock2018decision,hancock2021accumulation}. The lower panel of Figure \ref{fig:StandardModels} depicts how these models have previously been operationalised.
        
        However, as is the case for econometric choice models, these models have rarely been implemented with the incorporation of `process' data such as physiological data. All choice models, whether they are based on econometric, psychological or alternative foundations, are merely representations of the decision-making process, thus we do not observe the choice deliberation process. The key difference between models is that psychological models have `process' parameters that relate to how preferences for the alternatives change over time \textit{within a given choice task}. If we can inform these parameters with information relating to, for example, the perception or processing of information, then this may give psychological models a distinct advantage \citep{bansal2024discrete}. Figure \ref{fig:Process} demonstrates conceptually why such models may provide a psychologically appealing framework for the incorporation of such data, as new functions can be used to inform the process parameters based on physiological data representing some element of the choice deliberation process.
        
        \begin{figure}[ht!]
        	\centering
        	\caption{A conceptual framework for informing the `process parameters' within a psychological choice model.}
        	\label{fig:Process}	\includegraphics[scale=0.475]{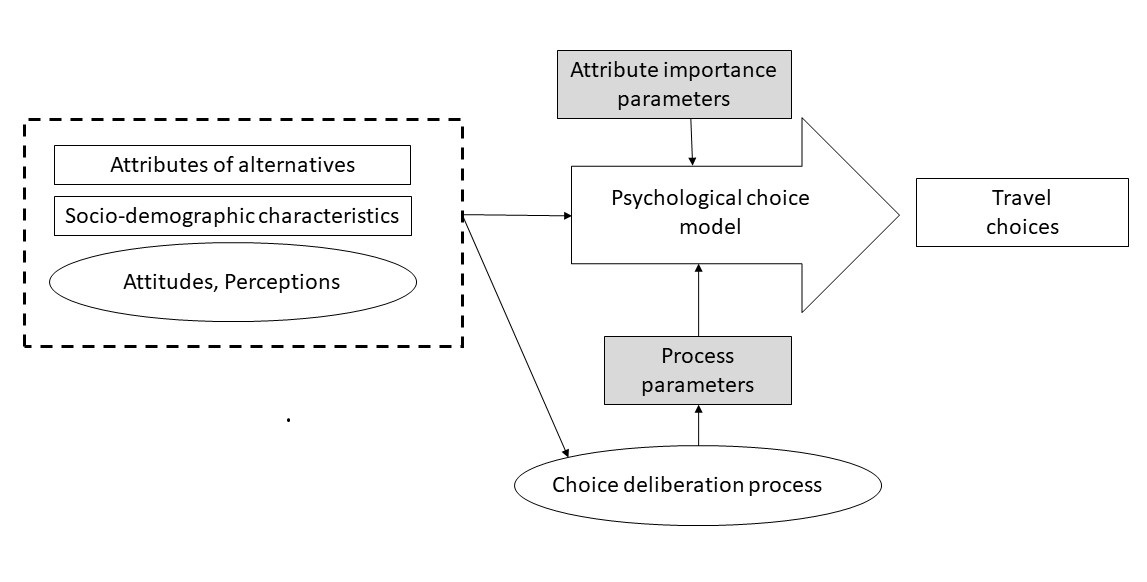}
        \end{figure}
        
        This leads us to our key research questions:
        (1) Are standard econometric choice models still sufficient if we also incorporate decision-making `process' (specifically eye-movement and stress measurement) data? (2) If not, does a move towards psychological choice models, where the choice deliberation process is `explicitly' modelled, provide a more favourable framework? 
        (3) Does having the new information inferred from the eye-tracking data lead to more behaviourally realistic models and/or better insights, and can this be demonstrated empirically?
        (4) Are the improvements (if any) gained from incorporating the eye-tracking data context dependent? 
        
        For investigating these research questions, we use both static and dynamic choice datasets to investigate whether decision field theory (DFT, a psychological choice model that has already been demonstrated to be effective for stated preference choice modelling \citealp{hancock2018decision,hancock2021accumulation}), relative to econometric choice models, is better suited to the inclusion of eye-tracking and stress information, and whether this leads to more accurate choice predictions and/or more interesting behavioural insights. To gain more comprehensive insights on the contributions of eye-tracking data to the different models and to investigate the extent of context-dependence, we specifically consider two very different choice contexts. In the first case, we consider the incorporation of eye-tracking data from a stated preference survey on accommodation choice. The second case study builds upon the work of \citet{paschalidis2018modelling}, who demonstrated that stress measurement data could be used to enhance econometric choice models for driving behaviour modelling. The present research takes this further by (a) testing whether psychological choice models are more suited for dynamic choice contexts and (b) leveraging eye-tracking data to get further insights on the decision-making process and the behaviour. 
        
        The remainder of this paper is organised as follows. First, we give an overview of psychological choice models and how they have been previously used. We specifically focus on decision field theory, describing the underlying mathematics, detailing how probabilities are generated from the model and discussing methods for the incorporation of physiological data. Second, we present our empirical applications, which cover both static and dynamic choice contexts: accommodation choice and gap acceptance at an unsignalised intersection. Third, we discuss conclusions and some next steps for future research in the context of choice modelling research.

        \section{Overview of psychological choice models}
        
        In a stream of research parallel but almost completely separate to econometric choice modelling, mathematical psychologists have developed choice models that have very different aims and objectives \citep{busemeyer2019cognitive}. These \textit{sequential sampling models} are often designed to `mimic' the decision-making process, with the key idea being that a decision-maker attends to different attributes/factors of the different alternatives over time within a single choice deliberation process, resulting in preference formation where the relative preference for the different alternatives increases and decreases over time until the decision-maker reaches a conclusion. An example of what this might look like is given in Figure \ref{fig:DFTpref}.
        
        \begin{figure}[ht!]
        	\centering
        	\caption{An example evolution of preferences during the course of deliberating over four alternatives, adapted from \citet{hancock2021accumulation}.}
        	\label{fig:DFTpref}
        	\includegraphics[scale=0.5]{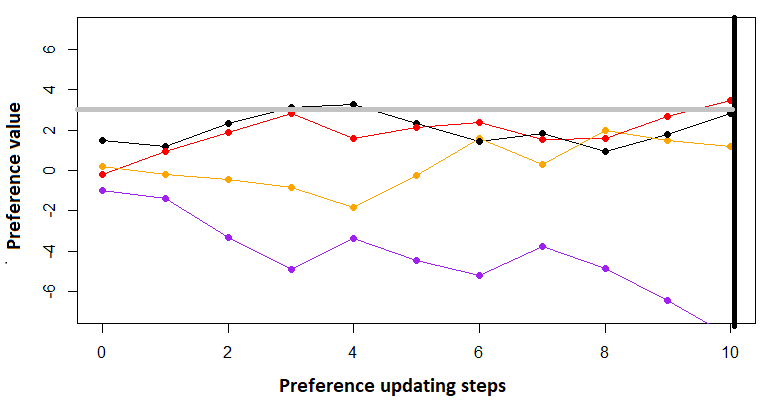}
        \end{figure}
            	
        The decision-maker may choose an alternative when the preference for an alternative reaches an \textit{internal threshold} (i.e. equivalent to satisficing as in \citealt{schwartz2002maximizing,gonzalez2018stochastic}, and represented by the horizontal grey line in Figure \ref{fig:DFTpref}). For example, a driver may decide that a gap is large enough and choose to merge onto a motorway from a slip road. Alternatively, they may continue to deliberate until they face some \textit{external threshold} such as a time limit upon which they cannot continue to deliberate (represented by the black vertical line after 10 preference updating steps in Figure \ref{fig:DFTpref}). For instance, at the point where a driver reaches the end of a slip road when merging onto a motorway, whereupon some merging strategies are no longer possible. The decision-maker may choose different alternatives depending on how long they consider the different alternatives. A shorter deliberation time, for example, may result in the decision-maker being more likely to choose their status quo option, for which there may be an initial bias. This is also represented in Figure \ref{fig:DFTpref}, with the alternative represented by black dots starting with a higher preference value and being chosen if the decision-maker concludes deliberating after 4 preference updating steps, but not 10. 
        
        There are a number of sequential sampling models that incorporate these ideas, including decision field theory \citep{busemeyer1992fundamental,busemeyer1993decision,roe2001multialternative}, the leaky competing accumulator \citep{usher2004loss}, the attentional drift diffusion model \citep{krajbich2012attentional} and decision by sampling \citep{noguchi2018multialternative}, with a detailed summary of these models and other psychological choice models given by \citet{busemeyer2019cognitive}. All of these models, like econometric models, have a number of input variables that result in an output of probabilities for different outcomes. However, the focus is typically on capturing some element of the decision-making process (e.g. the distribution of choice response time, \citealp{brown2008simplest}) rather than the accuracy of the prediction of outcomes.
        
        Though these models have also not often integrated physiological data, a key exception is for eye-tracking information, which has been used extensively in marketing research \citep{shi2013information,thomas2019gaze} and inspired models such as the attentional drift diffusion model \citep{krajbich2012attentional,fisher2017attentional,tavares2017attentional}. These models however have a greater focus on replicating eye movement patterns rather than recovering chosen alternatives. Whilst there has been some work specifically linking process data to the process parameters in psychological choice models \citep{glockner2012processing,nunez2017attention}, there have been no such applications in large-scale choice modelling contexts. 
        
        In this paper, we build upon previous work on decision field theory (DFT), which has recently made the transition from mathematical psychology to large-scale choice modelling work in travel behaviour research \citep{hancock2018decision,hancock2021accumulation}. Previous applications of the model within this context have not included process data, and as such have relied on careful implementation to ensure empirical identification \citep{hancock2021accumulation}, as there is no information that is used specifically to inform estimates for the process parameters. We overcome this limitation by using physiological sensor data to inform the process parameters within DFT models. In the following section, we give the mathematical outline for DFT, detailing how the physiological sensor data is incorporated into the DFT models.

        \section{Methodology}
            
            In this section, we first look at a summary of the existing implementation of the chosen psychological choice model, decision field theory (DFT), before discussing how eye-tracking information and stress indicator data can be incorporated into a DFT model and why it is a conceptually promising framework.
	
            \subsection{Outline of existing decision field theory framework}\label{txt:DFT:basic}        	
        	Decision field theory (which was used to simulate the preference updating process demonstrated in Figure \ref{fig:DFTpref}) is a dynamic, stochastic model, first developed by \citet{busemeyer1992fundamental,busemeyer1993decision}, adapted for multiple alternatives by \citet{roe2001multialternative} and further developed for implementation in a larger-scale choice modelling context by \citet{hancock2018decision,hancock2021accumulation}. The work in this paper is based largely on \citet{hancock2021accumulation}'s framework which importantly incorporates relative importance scaling parameters for different attributes, as well as `attention weights', meaning it is ideally suited for the incorporation of eye-tracking information. Furthermore, it has four additional `process parameters,' which may allow for alternative or better methods for incorporating the choice process data.
        	
        	As a starting point, it is assumed that a decision-maker, $n$, has some initial preference value for each alternative in a choice set $s$ of size $J$, contained within a preference vector  $\bm{P}_{n,s,0}$ (at time point $\tau = 0$). The decision-maker `deliberates' over the different attributes of the alternatives, resulting in increases and decreases for each element in $\bm{P}_{n,s,\tau}$. At each `preference updating step', it is further assumed that the decision-maker attends to a single attribute (e.g. cost), comparing that attribute across the different alternatives. This results in the preference vector updating according to:
        	\begin{linenomath*}
        		\begin{equation}\label{eq:DFT:nextP}
        		\bm{P}_{n,s,\tau+1} = S_{n,s} \cdot \bm{P}_{n,s,\tau}+\bm{V}_{n,s,\tau+1},
        		\end{equation}
        	\end{linenomath*}
        	where $\tau$ represents the number of preference updating steps, $S$ is a feedback matrix and $\bm{V}$ is a random valence vector which contains information on the attribute attended to in the current step. At some point, it is assumed that the decision-maker reaches a conclusion. For the work in this paper, we use \citet{hancock2021accumulation}'s framework for which DFT models have an external time threshold (see Figure \ref{fig:DFTpref}), meaning that the number of preference updating steps, $\tau$, is an estimated parameter. 
        	
        	The feedback matrix contains two more of the process parameters and is defined as:
        	\begin{linenomath*}
        		\begin{equation}\label{eq:feedback}
        		S_{n,s} = I_{n,s}-\phi_2 \times  \exp (-\phi_1 \times D_{n,s}^2),
        		\end{equation}
        	\end{linenomath*}
        	where $\phi_1$ is a sensitivity parameter, which impacts how much similar alternatives `compete' with each other during the deliberation process and thus controls the level of contextual effects predicted by the model \citep{roe2001multialternative}. $\phi_2$ is a memory parameter, which theoretically allows for differences in the relative importance of information considered at the start and end of a choice deliberation process, but which has never been mathematically used as such due to DFT having never been applied to complex choice data with the required information regarding the choice process. $I_{n,s}$ is an identity matrix of size $J \times J$, and $D_{n,s}$ is a matrix containing the Euclidean distances between the full set of pairs of alternatives measured with respect to the attribute levels and the relative importance of the different attributes (see \citealt{hancock2021accumulation}). 
        	
        	Next, we consider the random valence vector. At step $\tau$, we have $\bm{V}_{n,s,\tau}$, which can be calculated as:
        	\begin{linenomath*}
        		\begin{equation}\label{eq:DFT:Vt}
        		\bm{V}_{n,s,\tau} = C_{n,s} \cdot M_{n,s} \cdot \beta \cdot \bm{W}_{n,s,\tau}+ \bm{\varepsilon}_{n,s,\tau},
        		\end{equation}
        	\end{linenomath*}
        	where $C_{n,s}$ is a contrast matrix used to rescale the attribute differences such that they sum to zero (see \citealt{roe2001multialternative}). $M_{n,s}$ is the matrix of attribute values dependent on the specific choice task, with each element multiplied by a corresponding attribute scaling coefficient, contained in the diagonal matrix $\beta$. We also have $\bm{W}_{n,s,\tau}=[ 0 .. 1 .. 0 ]'$ with the $k^{th}$ entry, i.e. $\bm{W}_{n,s,\tau,k} =1$ if and only if attribute $x_k$ is the attribute being attended to by the decision-maker at preference updating step $\tau$. We thus have attribute attention weights, $w_k$, where $\sum_{k=1}^{K} w_k = 1$ ($K$ being the number of attributes) and $w_k$ represents the probability that the decision-maker attends to attribute $x_k$. The inclusion of both attention weights and scaling parameters for each attribute does not lead to significant gains for choice-data alone \citep{hancock2021accumulation}. However, it allows the model to capture not only attribute importance but information regarding how the decision is made. For example, the scaling parameters could capture the attribute importance (i.e. how important the attribute is in capturing preferences), whilst the attention weights could be directly linked to the choice deliberation process and thus become process parameters, informing how often a decision-maker visually attends a particular attribute. We thus have parameters regarding the perception of information, meaning we can create functions relating perception to the process parameters (represented graphically in Figure \ref{fig:Process}).
        	
        	The final process parameter is `process noise' entered through the variance of the error term, with $\bm{\varepsilon}_\tau=[ \varepsilon_1 .. \varepsilon_{J_{n}}]' $, and $\varepsilon_i \sim N(0,\sigma_\epsilon^2)$, distributed identically and independently across alternatives, steps, individuals and choice tasks. The result of having this error term is that there are two means for generating `noise' within a DFT model, as the random attribute attendance also results in probabilistic choices. 
        	
        	To estimate the probabilities of different alternatives being chosen under this DFT model, we require the expected value and the covariance of the preference values ($\bm{\xi}_{n,s,\tau}$ and $\Omega_{n,s,\tau}$), where
        	
        	\begin{linenomath*}
        		\begin{equation}\label{eq:ExpP}
        			E[\bm{P}_{n,s,\tau}] = \bm{\xi}_{n,s,\tau} = (I_{n,s}-S_{n,s})^{-1} (I_{n,s}-S_{n,s}^\tau) \cdot \bm{\mu}_{n,s} + S_{n,s}^\tau \cdot \bm{P}_{n,s,0},
        		\end{equation}
        	\end{linenomath*}
        	and
        	\begin{linenomath*}
        		\begin{equation}\label{eq:CovP}
        			Cov[\bm{P}_{n,s,\tau}]= \Omega_{n,s,\tau} = \sum_{r=0}^{\tau-1} \left[ S_{n,s}^{r} \cdot \Phi_{n,s} \cdot S_{n,s}^{r'} \right].
        		\end{equation}
        	\end{linenomath*}
        	$\bm{\mu}_{n,s}$ and $\Phi_{n,s}$ are the expectation and covariance of the valence vector, respectively, with full formulations and derivations for these and the above equations given by \citet{hancock2021accumulation}. This results in a choice probability for choosing alternative $j$ from the set $CS_{n,s}$ at step $\tau$ of:
        	\begin{linenomath*}
        		\begin{equation}\label{eq:DFTprob}
        		\begin{gathered}
        		Prob \left[  \bm{P}_{n,s,\tau} \left[j\right] = \max_{i \in CS_{n,s}}  \bm{P}_{n,s,\tau}\left[i\right] \right] =\\ \int_{\bm{X}_{n,s,\tau}>0} exp \left[ -(\bm{X}_{n,s,\tau}- \bm{\Gamma}_{n,s,\tau})' \Lambda_{n,s,\tau}^{-1} (\bm{X}_{n,s,\tau}- \bm{\Gamma}_{n,s,\tau})/2 \right] / (2 \pi |\Lambda_{n,s,\tau}|^{0.5}) dX.
        		\end{gathered}
        		\end{equation}
        	\end{linenomath*}
        	with full details of this calculation again given by \citet{hancock2021accumulation}.

            \subsection{The incorporation of eye-tracking data}
            
            Though eye-tracking data has been used recently to augment traditional choice models \citep{bansal2024discrete}, as far as we are aware, eye-tracking information has never been incorporated within a DFT model. Under the above specification of DFT, there are two parameters associated with each attribute. Whilst this does not result in confounding due to the different mathematical function of the two parameters, setting both as free estimated parameters does not result in a substantial improvement in model fit \citep{hancock2021accumulation}. In other words, there is not enough information in choice data alone for both of these parameters to be useful. However, if we additionally consider process data, it becomes possible that we may be able to clearly differentiate the purpose of these two parameters. Conceptually, the scaling parameters might capture attribute importance, whilst the attention weight refers to the percentage of time a decision-maker considers an attribute. Thus, if we have information regarding a decision-maker's eye fixations when considering the different alternatives, we could assume that fixating on an attribute directly corresponds to `considering' that attribute. However, the actual choice deliberation process is still unknown, with, for example, it being possible that a decision-maker considers different attributes or factors from the ones they are looking at, be distracted or have actually already made their decision. As such, we can estimate a parameter ($\alpha$) for the `relative importance' of gaze fixations (v), with the attention weight ($w_{k,n,s}$) for some attribute $k$ thus being based on some general function:
            \begin{linenomath*}
        		\begin{equation}\label{eq:weights1}
        		    w_{k,n,s} = f(z_n,\delta_s,\alpha,v,\epsilon),
        		\end{equation}
        	\end{linenomath*}
            such that the attribute attention weight is only partly based on how much visual attention it receives, and is also a function of individual characteristics, $z_n$, features of the choice task $\delta_s$ (such as choice task difficulty) and some error $\epsilon$. 
            
            An alternative conceptual framework would instead suggest that a better approach would be to link visual attention with the attribute scaling parameters, as studies suggest that decision-makers spend longer looking at more important (or more favourable) attributes or alternatives \citep{shimojo2003gaze}. A scaling parameter $s_k$ for attribute $k$ could thus equivalently be defined as a function of individual characteristics, choice task features and visual attention. This approach is of course also possible within a standard econometric modelling framework if marginal utility parameters are linked to visual attention.
            
            In our empirical work, we consider both possible frameworks, through basic functions that link attribute attention weights and scaling functions to visual attention. 
            
            \subsection{The incorporation of stress indicator data}
        
            Despite the fact that decision field theory was originally developed partly to explain decision-making under time pressure and can do so effectively \citep{diederich1997dynamic,pettibone2012testing}, it has never been applied with process data relating to how stressed the decision-maker actually is. There are many types of physiological data that can be recorded to estimate stress, including but not limited to heart rate, electrodermal activity, blood volume pulse and temperature. Thus, as is the case for eye-tracking data, there are a number of possible inputs or types of information that can be used. This also means that there are many methods for incorporating the stress level ($\alpha_{n,s}$) of individual $n$ in choice task $s$ within the choice model. We consider two distinct possibilities for the DFT model. 
            
            The first, `simple' option, is to include stress indicators in the initial preference values. This is a conceptually appealing possibility as a decision-maker who is stressed may be predisposed not to make a particular choice. For example, a driver in a hurry will likely accept smaller gap sizes when merging from a slip road onto a motorway. This also mirrors methods for including stress data in econometric choice models, with stress indicators being effectively added to utility by \citet{paschalidis2018modelling}. 
            
            A second alternative method for incorporating stress indicators in a DFT model is to instead reparameterise one or more of DFT's process parameters as a function of the stress measurement data. For example, the process noise ($\sigma_{\epsilon}$) within a DFT model could increase when a decision-maker is stressed and/or distracted and not properly focused, with driving behaviour studies demonstrating that drivers make different decisions when they are stressed \citep{rendon2016effects,paschalidis2019developing} or distracted \citep{hurts2011distracted,foss2014distracted,li2018does}. We study both of these options in our empirical application on driving behaviour.

            \section{Empirical application: static choice scenario (accomodation choice)}
            
            For our empirical applications, we trial our theories by first testing the addition of physiological information to static choice scenarios, where decision-makers consider a number of traditional stated preference choice tasks. Since the attribute values corresponding to each choice are unchanged throughout the decision-making process, the  key aspect for understanding which alternatives are chosen is to understand which attributes are perceived to be important and have the greatest influence on preferences. For the work in this paper, it is assumed that the relative time spent looking at the different attributes indicates the relative importance of the different attributes. We next present the dataset, outline the modelling framework, present the results of the models without (referred as 'Base model') and with eye-tracking data. In both types of model, the performance of the multinomial logit (MNL) and DFT based frameworks have been investigated.
            
                \subsection{Dataset}\label{sec:StaticData}
                For our static choice scenarios, we consider a stated preference dataset collected by \citet{cohen2017multi}. 38 participants completed 45 stated preference tasks where in each case they had to choose the best apartment from a set of three. Each apartment had a 1-5 star rating for the attributes: ease of transportation, size, condition and kitchen facilities. An example choice scenario is displayed in Figure \ref{fig:AccChoice}.
                
                \begin{figure}[ht!]
                	\centering
        	        \caption{An example choice task for the SP dataset \citep{cohen2017multi}.}
        	        \label{fig:AccChoice}
        	        \includegraphics[scale=0.75]{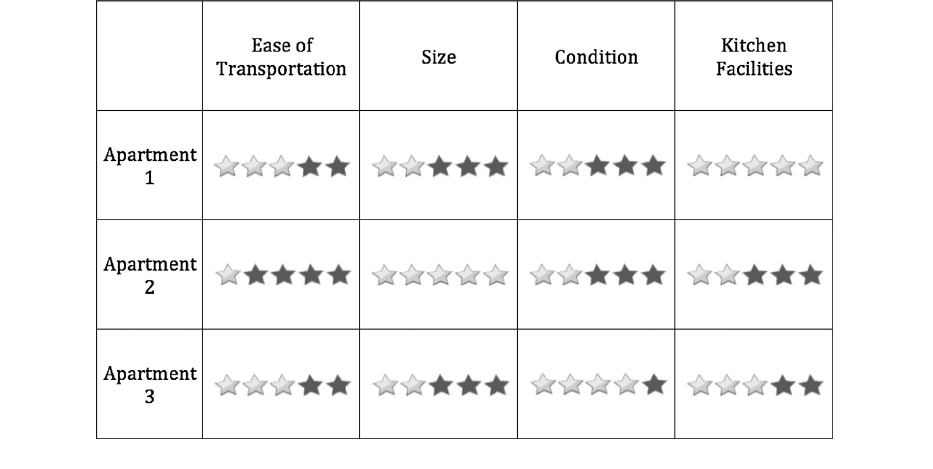}
                \end{figure}
                
                For each participant, there were five `catch trials' with a dominant alternative. 6 decision-makers failed at least one catch trial and were removed from the dataset. Further data cleaning, detailed by \citet{cohen2017multi} resulted in a total of 1,430 responses remaining for analysis. Eye-tracking information was collected using EyeLink 1000 (SR Research), with a drift correction procedure being used before participants completed each choice task. Gaze fixations were determined using Data Viewer (SR Research) and could thus be assigned to a particular alternative and attribute, with the fixation duration also being recorded. This meant that we could generate two key variables for each attribute for each choice task: a fixation count share and a viewing time share. This information can then be used to inform the relative importance of each attribute within our DFT models.
                
                \subsection{Base model}
                
                \subsubsection{Model specifications}
                
                We test our DFT models against multinomial logit models (MNL). Our base MNL model only uses four marginal utility parameters, one for each attribute, such that the utility for alternative $j$ for decision-maker $n$ in choice task $s$ is:
                
                 \begin{linenomath*}
            	    \begin{equation}\label{eq:MNL_static1}
            		    U_{n,s,j}= \beta_{Kitchen} \cdot x_{K_{n,s,j}} + \beta_{Condition} \cdot x_{C_{n,s,j}}  + \beta_{Size} \cdot x_{S_{n,s,j}} + \beta_{Transport} \cdot x_{T_{n,s,j}} +\epsilon_{n,s,j},
            		\end{equation}
            	\end{linenomath*}
                
                where the $x_K$, $x_C$, $x_S$ and $x_T$ are the attributes values for kitchen facilities, condition, size and ease of transportation, respectively. Alternative specific constants for each alternative are not included as they are not found to be significant during testing of the models. The assumption of type I extreme value error terms for $\epsilon$  results in probabilities of choosing each alternative being generated using the well known MNL formula:
    
                \begin{linenomath}
                   \begin{equation}\label{Eq:MNL_prob}
                      P_{n,s,j} = \frac{exp(U_{n,s,j})}{\sum_{i=1}^{3} exp(U_{n,s,i})}.
                   \end{equation}
                \end{linenomath}
                
                For our first DFT model, we estimate attention weights for each attribute, such that we generate $\bm{w}$, the vector for the probability of attending each attribute (which is the expectation of $W_{n,s,\tau}$, used in the calculation of $\mu_{n,s}$, which in turn is used to calculate probabilities of choosing the different alternatives):
                \begin{linenomath}
                   \begin{equation}\label{Eq:DFT_weight}
                      \bm{w}  = \begin{bmatrix}
            		    w_{Kitchen}\\
            		    w_{Condition}\\
            		    w_{Size} \\
            		    w_{Transport}
            		    \end{bmatrix}.
                   \end{equation}
                \end{linenomath}
                To ensure that these four elements sum to 1, we use logistic transformations such that, for example:
                \begin{linenomath}
                   \begin{equation}\label{Eq:staticDFT1}
                      w_{Kitchen} = \frac{exp(\gamma_{Kitchen})}{exp(\gamma_{Kitchen})  + exp(\gamma_{Condition})  + exp(\gamma_{Size})  + exp(\gamma_{Transport})},
                   \end{equation}
                \end{linenomath}
                where fixing one $\gamma$ to zero ensures model identification. 
                For our second DFT model, we fix the attention weight parameters to be of equal value and instead estimate attribute scaling parameters \citep{hancock2021accumulation}, which allow for the relative importance of the attributes to be estimated whilst assuming that each attribute is considered equally often during the choice deliberation process. We thus have a set of scaling factors:  
                
                \begin{linenomath*}
            	    \begin{equation}\label{eq:staticDFT2}
            		    \bm{\beta}  = \begin{bmatrix}
            		    \beta_{Kitchen}\\
            		    \beta_{Condition}\\
            		    \beta_{Size} \\
            		    \beta_{Transport}
            		    \end{bmatrix},
            		\end{equation}
            	\end{linenomath*}
                which are estimated with $\beta$-coefficients, equivalently to our MNL model. Our DFT models additionally have four `process' parameters to be estimated. Two of these parameters, however, are fixed.\footnote{Whilst we could fix the error term in our DFT model which estimates scaling parameters, such that the model `matches' the MNL model in having four estimated $\beta$-coefficients, we instead choose to fix the first $\beta$ in our DFT model such that it more closely matches the DFT model with estimated attention weights, as this model requires estimation to obtain the size of the error term.} Firstly, initial model testing found a very high estimate for the number of preference updating steps. This can cause estimation issues, as there is no change in the preference values between preference updating steps once they have stabilised \citep{berkowitsch2014rigorously}, and thus a higher estimate for the number of preference accumulation steps results in no change in model results. We thus fix the number of preference updating steps, $\tau = 1000$. Secondly, if $\phi_2 > 1/J$, where $J$ is the number of alternatives, then the structure of the feedback matrix (Equation \ref{eq:feedback}) can result in the creation of non-invertible matrices, which results in it not being possible to calculate the probabilities of the different alternatives (see Equation \ref{eq:ExpP}). However, using a logistic transformation to estimate $\phi_2$ with an upper limit of $1/3$ ($J=3$ in this case) results in extremely high estimates ($\phi_2$ tends towards 1/3). Thus, we also fix $\phi_2=1/3$. Our DFT models, equivalent to the MNL models, find no baseline preferences towards the different alternatives, thus parameters for estimating the size of biases (similar to alternative-specific constants in the MNL models) are not included. As a result, the DFT models have one extra estimated parameter in comparison to the MNL models ($\phi_1$ and $\sigma_\epsilon$ are estimated, but there is one less parameter required to estimate the relative importance of the attributes).     
                
                \subsubsection{Results}
                
                The results of the base MNL and two base DFT models are given in Table \ref{tab:static1}. DFT-B1 estimates the attention weight parameters rather than the scaling parameters for the relative importance of the different attributes. Conversely, DFT-B2 estimates only scaling parameters.
                
                \begin{table}[ht!]
                  \centering
                  \scriptsize
                  \caption{Base model results without incorporating eye-tracking information for the static stated preference data.}
                 \begin{tabular}{|c|rr|cc|cc|}
                    \toprule
                    Model & \multicolumn{2}{c|}{MNL-B} & \multicolumn{2}{c|}{DFT-B1} & \multicolumn{2}{c|}{DFT-B2} \\
                    \midrule
                    Description & \multicolumn{6}{c|}{Base models with no eye-tracking information included} \\
                    \midrule
                    Log-likelihood(null) & \multicolumn{6}{c|}{-1,571.02} \\
                    \midrule
                    Log-likelihood & \multicolumn{2}{c|}{-1,329.09} & \multicolumn{2}{c|}{-1,320.56} & \multicolumn{2}{c|}{-1,322.76} \\
                    Estimated parameters & \multicolumn{2}{c|}{4} & \multicolumn{2}{c|}{5} & \multicolumn{2}{c|}{5} \\
                    Adj. $\rho^2$ & \multicolumn{2}{c|}{0.1514} & \multicolumn{2}{c|}{0.1562} & \multicolumn{2}{c|}{0.1548} \\
                    BIC   & \multicolumn{2}{c|}{2,687.23} & \multicolumn{2}{c|}{2,677.46} & \multicolumn{2}{c|}{2,681.85} \\
                    \midrule
                          & \multicolumn{1}{c}{est.} & \multicolumn{1}{c|}{rob. t-rat.} & est.  & rob. t-rat. & est.  & rob. t-rat. \\
                    \midrule
                    $\beta_{Kitchen}$ & \multicolumn{1}{c}{0.42} & \multicolumn{1}{c|}{6.75} & 1.00  & \textbf{fixed} & 1.00  & \textbf{fixed} \\
                    $\beta_{Condition}$ & \multicolumn{1}{c}{0.93} & \multicolumn{1}{c|}{13.41} & 1.00  & \textbf{fixed} & 2.42  & 7.05 \\
                    $\beta_{Size}$ & \multicolumn{1}{c}{0.60} & \multicolumn{1}{c|}{9.43} & 1.00  & \textbf{fixed} & 1.44  & 7.06 \\
                    $\beta_{Transport}$ & \multicolumn{1}{c}{0.57} & \multicolumn{1}{c|}{10.21} & 1.00  & \textbf{fixed} & 1.38  & 6.00 \\
                    \midrule
                    $\gamma_{Kitchen}$ & \cellcolor[rgb]{ .851,  .851,  .851} & \cellcolor[rgb]{ .851,  .851,  .851} & 0.00  & \textbf{fixed} & 0.00  & \textbf{fixed} \\
                    $\gamma_{Condition}$ & \cellcolor[rgb]{ .851,  .851,  .851} & \cellcolor[rgb]{ .851,  .851,  .851} & 0.88  & 5.76  & 0.00  & \textbf{fixed} \\
                    $\gamma_{Size}$ & \cellcolor[rgb]{ .851,  .851,  .851} & \cellcolor[rgb]{ .851,  .851,  .851} & 0.37  & 2.48  & 0.00  & \textbf{fixed} \\
                    $\gamma_{Transport}$ & \cellcolor[rgb]{ .851,  .851,  .851} & \cellcolor[rgb]{ .851,  .851,  .851} & 0.30  & 1.71  & 0.00  & \textbf{fixed} \\
                    \midrule
                    $\phi_1$ & \cellcolor[rgb]{ .851,  .851,  .851} & \cellcolor[rgb]{ .851,  .851,  .851} & 1.19E-04 & 3.13  & 3.04E-05 & 3.13 \\
                    $\phi_2$ & \cellcolor[rgb]{ .851,  .851,  .851} & \cellcolor[rgb]{ .851,  .851,  .851} & 0.33  & \textbf{fixed} & 0.33  & \textbf{fixed} \\
                    $\sigma_{\epsilon}$ & \cellcolor[rgb]{ .851,  .851,  .851} & \cellcolor[rgb]{ .851,  .851,  .851} & 25.03 & 14.17 & 38.39 & 6.58 \\
                    $\tau$ & \cellcolor[rgb]{ .851,  .851,  .851} & \cellcolor[rgb]{ .851,  .851,  .851} & 1,000.00 & \textbf{fixed} & 1,000.00 & \textbf{fixed} \\
                    \bottomrule
                  \end{tabular}%
                  \label{tab:static1}%
                \end{table}%
                
                We observe a better model fit for our DFT models compared to the MNL model, with the DFT models returning better adjusted $\rho^2$ and BIC values. DFT-B1 (with estimated attention weights) outperforms DFT-B2 (with estimated scaling coefficients). This result is in contrast to the findings of \citet{hancock2021accumulation}, who found that DFT models with estimated scaling parameters consistently outperformed DFT models with estimated attention weights. One possible cause for this is that \citet{hancock2021accumulation} tested their models on data with large scaling differences between the attributes. In this case, with the exception of condition, the $\beta$ or $\gamma$ estimates are not substantially different from each other within the MNL model or within the DFT models. The $\gamma$ estimates of DFT-B1, once exponentiated (1.00, 2.40, 1.45, 1.35), are remarkably similar to those of DFT-2, suggesting that the 2 units gain in log-likelihood is a direct result of the impact non-equal attention weights has on the covariance of the preferences (see Equation \ref{eq:CovP}). MNL-B also finds similar coefficients (which, once adjusted for scale, are 1.00, 2.20, 1.42 and 1.36) to the DFT models, though this implies that the DFT models assign more importance to the condition of the apartment. 
    
                \subsection{Models incorporating eye-tracking information}
                
                \subsubsection{Model framework} 
                
                The eye-tracking information (detailed in Section \ref{sec:StaticData}) gives us two key variables for each attribute in each choice task. We thus have, for example, $x_{gazecount,K_{n,s}}$ and $x_{gazetime,K_{n,s}}$, which represent the percentage of fixations that are on the ratings of the kitchen facilities and the percentage of total choice deliberation time spent fixating on these ratings (both summed across the ratings of all alternatives). We have two distinct methods for how to include these new variables in a DFT model.\footnote{Note that future applications may alternatively use a latent variable approach to account for possible measurement error.} 
                
                The first possibility for the DFT model is the psychologically more `intuitive' option of adjusting the attention weight parameters depending on how much time is spent fixating on the different attributes. We thus adjust Equation \ref{Eq:staticDFT1} such that eye-tracking information is incorporated by replacing $\gamma_{Kitchen}$ with:
                \begin{linenomath}
                   \begin{equation}\label{Eq:DFT_weight2}
                      \gamma_{Kitchen} + x_{gazecount,K_{n,s}} \cdot \alpha_{gaze_{count}} + x_{gazetime,K_{n,s}} \cdot \alpha_{gaze_{time}},  
                   \end{equation}
                \end{linenomath}
                where $\alpha_{gaze_{count}}$ and $\alpha_{gaze_{time}}$ are the relative importance
                of fixation counts and total times, respectively. Equivalent adjustments are made for the other three $\gamma$ terms. By adjusting the $\gamma$ terms, we obtain a separate $\bm{w}_{n,s}$ for each individual $n$ in each choice task $s$ that is adjusted depending on how much the decision-maker looks at each attribute in the choice task in question, but still has the key property that the four attention weights sum to one across all $n$ and $s$. This specification allows the attention weights to depend on the eye-tracking data such that they actually represent `attention' rather than solely being used to capture preference. 
                
                Alternatively, the second option is to instead adjust the scaling terms, $\beta$, instead of $\gamma$. This method can also be used to incorporate eye-tracking information in the MNL models, and implies that the relative importance of the attributes depends on how much the individual looks at them. In of itself, this is a reasonable assumption, but having attention weights as well as scaling parameters allows for the possibility that a decision-maker takes more value or more meaning from some attributes, but may still look at attributes that are not of importance to them. 
                
                A further complexity arises in that either base DFT model can be used in each case, giving four different parameterisations of the DFT model, as the preference and attention for attributes can both be captured by either of the attention or scaling parameters. All four possibilities are tested in the subsequent section.

                \subsubsection{Results}
            
                Before we test the parameterisation of where to include eye-tracking information in a DFT model, we first test the relative influence of the two different variables that can be included in our models: the fixation count share and the viewing time share. We test 3 DFT models, all incorporating eye-tracking information into the attention weights and estimating relative importance weights as a base. The first model includes only count share, the second only viewing time share, and the third includes both. The results of these three models are given in Table \ref{tab:static2}.
                
                \begin{table}[ht!]
                  \centering
                  \scriptsize
                  \caption{The results of incorporating fixation count and viewing time shares into the attribute attention weights in DFT models.}
                 \begin{tabular}{|c|cc|cc|cc|}
                    \toprule
                    Model & \multicolumn{2}{c|}{DFT-E1} & \multicolumn{2}{c|}{DFT-E2} & \multicolumn{2}{c|}{DFT-E3} \\
                    \midrule
                    Base model & \multicolumn{6}{c|}{DFT-B2} \\
                    \midrule
                    Additional information & \multicolumn{2}{c|}{fixation counts} & \multicolumn{2}{c|}{fixation times} & \multicolumn{2}{c|}{both counts and times} \\
                    \midrule
                    location of $\alpha$ pars. & \multicolumn{6}{c|}{$\gamma$} \\
                    \midrule
                    Log-likelihood(null) & \multicolumn{6}{c|}{-1,571.02} \\
                    \midrule
                    Log-likelihood & \multicolumn{2}{c|}{-1,236.50} & \multicolumn{2}{c|}{-1,245.91} & \multicolumn{2}{c|}{-1,235.82} \\
                    Estimated parameters & \multicolumn{2}{c|}{6} & \multicolumn{2}{c|}{6} & \multicolumn{2}{c|}{7} \\
                    Improvement over DFT-B2 & \multicolumn{2}{c|}{86.26} & \multicolumn{2}{c|}{76.85} & \multicolumn{2}{c|}{86.94} \\
                    Adj. $\rho^2$ & \multicolumn{2}{c|}{0.2091} & \multicolumn{2}{c|}{0.2031} & \multicolumn{2}{c|}{0.2089} \\
                    BIC   & \multicolumn{2}{c|}{2,516.59} & \multicolumn{2}{c|}{2,535.41} & \multicolumn{2}{c|}{2,522.50} \\
                    \midrule
                          & est.  & rob. t-rat. & est.  & rob. t-rat. & est.  & rob. t-rat. \\
                    \midrule
                    $\beta_{Kitchen}$ & 1.00  & \textbf{fixed} & 1.00  & \textbf{fixed} & 1.00  & \textbf{fixed} \\
                    $\beta_{Condition}$ & 2.03  & 8.50  & 2.07  & 8.59  & 2.03  & 8.53 \\
                    $\beta_{Size}$ & 1.35  & 7.96  & 1.36  & 7.93  & 1.35  & 7.94 \\
                    $\beta_{Transport}$ & 1.27  & 6.95  & 1.27  & 6.96  & 1.27  & 6.96 \\
                    \midrule
                    $\gamma_{Kitchen}$ & 0.00  & \textbf{fixed} & 0.00  & \textbf{fixed} & 0.00  & \textbf{fixed} \\
                    $\gamma_{Condition}$ & 0.00  & \textbf{fixed} & 0.00  & \textbf{fixed} & 0.00  & \textbf{fixed} \\
                    $\gamma_{Size}$ & 0.00  & \textbf{fixed} & 0.00  & \textbf{fixed} & 0.00  & \textbf{fixed} \\
                    $\gamma_{Transport}$ & 0.00  & \textbf{fixed} & 0.00  & \textbf{fixed} & 0.00  & \textbf{fixed} \\
                    \midrule
                    $\phi_1$ & 1.30E-05 & 3.55  & 1.28E-05 & 3.39  & 4.60E-05 & 3.53 \\
                    $\phi_2$ & 0.33  & \textbf{fixed} & 0.33  & \textbf{fixed} & 0.33  & \textbf{fixed} \\
                    $\sigma_{\epsilon}$ & 38.39 & 6.58  & 32.71 & 7.87  & 32.03 & 7.83 \\
                    $\tau$ & 1,000.00 & \textbf{fixed} & 1,000.00 & \textbf{fixed} & 1,000.00 & \textbf{fixed} \\
                    \midrule
                    $\alpha_{gaze_{count}}$ & 2.29  & 7.38  & 0.00  & \textbf{fixed} & 1.86  & 4.63 \\
                    $\alpha_{gaze_{time}}$ & 0.00  & \textbf{fixed} & 1.92  & 6.36  & 0.43  & 0.97 \\
                    \bottomrule
                  \end{tabular}%
                  \label{tab:static2}%
                \end{table}%

            The first point to note is that all three models show a substantial improvement over the base model, DFT-B2. Gains of 86 and 77 log-likelihood units, respectively, are achieved for DFT-E1 and DFT-E2, despite these models having only a single additional parameter over DFT-B2. DFT-E3 does not result in a further improvement over these models, suggesting that $\alpha_{gaze_{count}}$ and $\alpha_{gaze_{time}}$  capture the same component in the model.\footnote{The same results are observed in MNL models, which are thus are not included, to avoid overcomplicating Table \ref{eq:staticDFT2}.}  As $\alpha_{gaze_{time}}$ has an insignificant coefficient in DFT-E3, we drop this parameter for the next set of models.
                
            We next consider the different parameterisations for how to incorporate $\alpha_{gaze_{count}}$. For MNL, we simply adjust the marginal utility parameters $\beta$. For DFT, we have four possibilities. We can adjust either the attention weights or the scaling parameters, and we can additionally estimate the constants for the attention weights or the scaling parameters (thus the models are built upon either DFT-B1 or DFT-B2). The results for all of these models are given in Table \ref{tab:static3}.
                
                \begin{table}[ht!]
                  \centering
                  \tiny
                  \caption{The results of different DFT models and the MNL model incorporating information on the relative share of gaze fixations for each attribute.} \begin{tabular}{|c|cc|cc|cc|cc|cc|}
                    \toprule
                    Model & \multicolumn{2}{c|}{MNL-E} & \multicolumn{2}{c|}{DFT-E1} & \multicolumn{2}{c|}{DFT-E4} & \multicolumn{2}{c|}{DFT-E5} & \multicolumn{2}{c|}{DFT-E6} \\
                    \midrule
                    Base model & \multicolumn{2}{c|}{MNL-B} & \multicolumn{2}{c|}{DFT-B2} & \multicolumn{2}{c|}{DFT-B1} & \multicolumn{2}{c|}{DFT-B2} & \multicolumn{2}{c|}{DFT-B1} \\
                    \midrule
                    Additional information & \multicolumn{10}{c|}{fixation counts, with a relative importance estimated by  $\alpha$} \\
                    \midrule
                    location of $\alpha$ pars. & \multicolumn{2}{c|}{$\beta$} & \multicolumn{2}{c|}{$\gamma$} & \multicolumn{2}{c|}{$\gamma$} & \multicolumn{2}{c|}{$\beta$} & \multicolumn{2}{c|}{$\beta$} \\
                    \midrule
                    Log-likelihood(null) & \multicolumn{10}{c|}{-1,571.02} \\
                    \midrule
                    Log-likelihood & \multicolumn{2}{c|}{-1,240.49} & \multicolumn{2}{c|}{-1,236.50} & \multicolumn{2}{c|}{-1,233.60} & \multicolumn{2}{c|}{-1,229.68} & \multicolumn{2}{c|}{-1,229.67} \\
                    Estimated parameters & \multicolumn{2}{c|}{5} & \multicolumn{2}{c|}{6} & \multicolumn{2}{c|}{6} & \multicolumn{2}{c|}{6} & \multicolumn{2}{c|}{6} \\
                    Improvement over base model & \multicolumn{2}{c|}{88.60} & \multicolumn{2}{c|}{86.26} & \multicolumn{2}{c|}{86.96} & \multicolumn{2}{c|}{93.08} & \multicolumn{2}{c|}{90.89} \\
                    Adj. $\rho^2$ & \multicolumn{2}{c|}{0.2072} & \multicolumn{2}{c|}{0.2091} & \multicolumn{2}{c|}{0.2110} & \multicolumn{2}{c|}{0.2134} & \multicolumn{2}{c|}{0.2135} \\
                    BIC   & \multicolumn{2}{c|}{2,517.30} & \multicolumn{2}{c|}{2,516.59} & \multicolumn{2}{c|}{2,510.80} & \multicolumn{2}{c|}{2,502.96} & \multicolumn{2}{c|}{2,502.93} \\
                    \midrule
                          & est.  & rob. t-rat. & est.  & rob. t-rat. & est.  & rob. t-rat. & est.  & rob. t-rat. & est.  & rob. t-rat. \\
                    \midrule
                    $\beta_{Kitchen}$ & 0.17  & \textbf{2.26} & 1.00  & \textbf{fixed} & 1.00  & \textbf{fixed} & 1.00  & \textbf{fixed} & 1.00  & \textbf{fixed} \\
                    $\beta_{Condition}$ & 0.62  & 7.21  & 2.03  & 8.50  & 1.00  & \textbf{fixed} & 5.12  & 2.19  & 1.00  & \textbf{fixed} \\
                    $\beta_{Size}$ & 0.31  & 4.51  & 1.35  & 7.96  & 1.00  & \textbf{fixed} & 2.10  & 2.30  & 1.00  & \textbf{fixed} \\
                    $\beta_{Transport}$ & 0.29  & 4.10  & 1.27  & 6.95  & 1.00  & \textbf{fixed} & 1.94  & 2.11  & 1.00  & \textbf{fixed} \\
                    \midrule
                    $\gamma_{Kitchen}$ & \cellcolor[rgb]{ .851,  .851,  .851} & \cellcolor[rgb]{ .851,  .851,  .851} & 0.00  & \textbf{fixed} & 0.00  & \textbf{fixed} & 0.00  & \textbf{fixed} & 0.00  & \textbf{fixed} \\
                    $\gamma_{Condition}$ & \cellcolor[rgb]{ .851,  .851,  .851} & \cellcolor[rgb]{ .851,  .851,  .851} & 0.00  & \textbf{fixed} & 0.72  & 5.71  & 0.00  & \textbf{fixed} & 0.68  & 5.68 \\
                    $\gamma_{Size}$ & \cellcolor[rgb]{ .851,  .851,  .851} & \cellcolor[rgb]{ .851,  .851,  .851} & 0.00  & \textbf{fixed} & 0.31  & 2.35  & 0.00  & \textbf{fixed} & 0.29  & 2.28 \\
                    $\gamma_{Transport}$ & \cellcolor[rgb]{ .851,  .851,  .851} & \cellcolor[rgb]{ .851,  .851,  .851} & 0.00  & \textbf{fixed} & 0.23  & 1.52  & 0.00  & \textbf{fixed} & 0.21  & 1.46 \\
                    \midrule
                    $\phi_1$ & \cellcolor[rgb]{ .851,  .851,  .851} & \cellcolor[rgb]{ .851,  .851,  .851} & 4.62E-05 & 3.55  & 1.28E-04 & 3.56  & 2.28E-06 & 0.97  & 2.03E-05 & 2.30 \\
                    $\phi_2$ & \cellcolor[rgb]{ .851,  .851,  .851} & \cellcolor[rgb]{ .851,  .851,  .851} & 0.33  & \textbf{fixed} & 0.33  & \textbf{fixed} & 0.33  & \textbf{fixed} & 0.33  & \textbf{fixed} \\
                    $\sigma_{\epsilon}$ & \cellcolor[rgb]{ .851,  .851,  .851} & \cellcolor[rgb]{ .851,  .851,  .851} & 38.39 & 6.58  & 22.72 & 15.87 & 125.56 & 1.90  & 51.37 & 4.66 \\
                    $\tau$ & \cellcolor[rgb]{ .851,  .851,  .851} & \cellcolor[rgb]{ .851,  .851,  .851} & 1,000.00 & \textbf{fixed} & 1,000.00 & \textbf{fixed} & 1,000.00 & \textbf{fixed} & 1,000.00 & \textbf{fixed} \\
                    \midrule
                    $\alpha_{gaze_{count}}$ & 1.59  & 7.82  & 2.29  & 7.38  & 2.41  & 7.08  & 14.55 & 1.65  & 6.00  & 3.17 \\
                    $\alpha_{gaze_{time}}$ & 0.00  & \textbf{fixed} & 0.00  & \textbf{fixed} & 0.00  & \textbf{fixed} & 0.00  & \textbf{fixed} & 0.00  & \textbf{fixed} \\
                    \bottomrule
                  \end{tabular}%
                  \label{tab:static3}%
                \end{table}%

                Similarly to the DFT models, MNL-E records a substantial improvement over MNL-B. All the DFT models however outperform the MNL-E model. Notably, it is the models with eye-tracking information included on the scaling parameters (DFT-E5 and DFT-E6) that record the best model fit, with greater improvements over the base models than MNL-E achieves over MNL-B. 
                
                As a contrast to previous results, where estimating constants for the attention weights (DFT-B1) resulted in better performance than estimating constants for the scaling parameters (DFT-B2), there is no difference between whether attention weight constants or scaling parameter constants are used, with DFT-E5 and DFT-E6 recording almost identical log-likelihoods. Despite this and the fact that the average difference in the probability of the chosen alternative under these two models is just 0.07\%, there is 3.83\% standard deviation between these probabilities. This is demonstrated in the left hand side of Figure \ref{fig:StaticComp}, which shows that DFT-E5 and DFT-E6 actually generate some large differences in probabilities for some choice tasks, though
                their total log-likelihoods are exactly the same.

                The right hand side of  Figure \ref{fig:StaticComp} demonstrates that larger differences between models are generated by including the eye-tracking information in the first place. DFT-B2, which does not include eye-tracking information, generates a smaller set of possible probabilities for the chosen alternatives as all individuals face the same set of choice tasks (represented in the figure by horizontal lines of points). DFT-E5 includes eye-tracking information, which results in a greater dispersion of probabilities for the different alternatives. 
                
                The positive estimates for $\alpha$ in DFT-E5 indicates that the longer the decision-maker spends looking at an attribute, the more that attribute will influence the probabilities of choosing the different alternatives. Consequently, the more that a decision-maker looks at attributes that favour their chosen alternative, the greater the probability of that alternative under the DFT model. An illustration of this is given by the red points (in Figure \ref{fig:StaticComp}), which denote chosen alternatives where more fixations than average (more than 25\%) were on an attribute with a level of 5 stars for the chosen alternative. This drives the substantial improvement in model fit that DFT-E5 achieves over DFT-B2, with, for example, a number of chosen alternatives having an estimated probability of more than 80\% under DFT-E5, but none doing so under DFT-B2.

                \begin{figure}[ht!]
            		\centering
            		\caption{The probability of the observed choices under DFT-E5 against DFT-E6, and DFT-E5 against DFT-B2, respectively.   }
            		\label{fig:StaticComp}\includegraphics[scale=0.35]{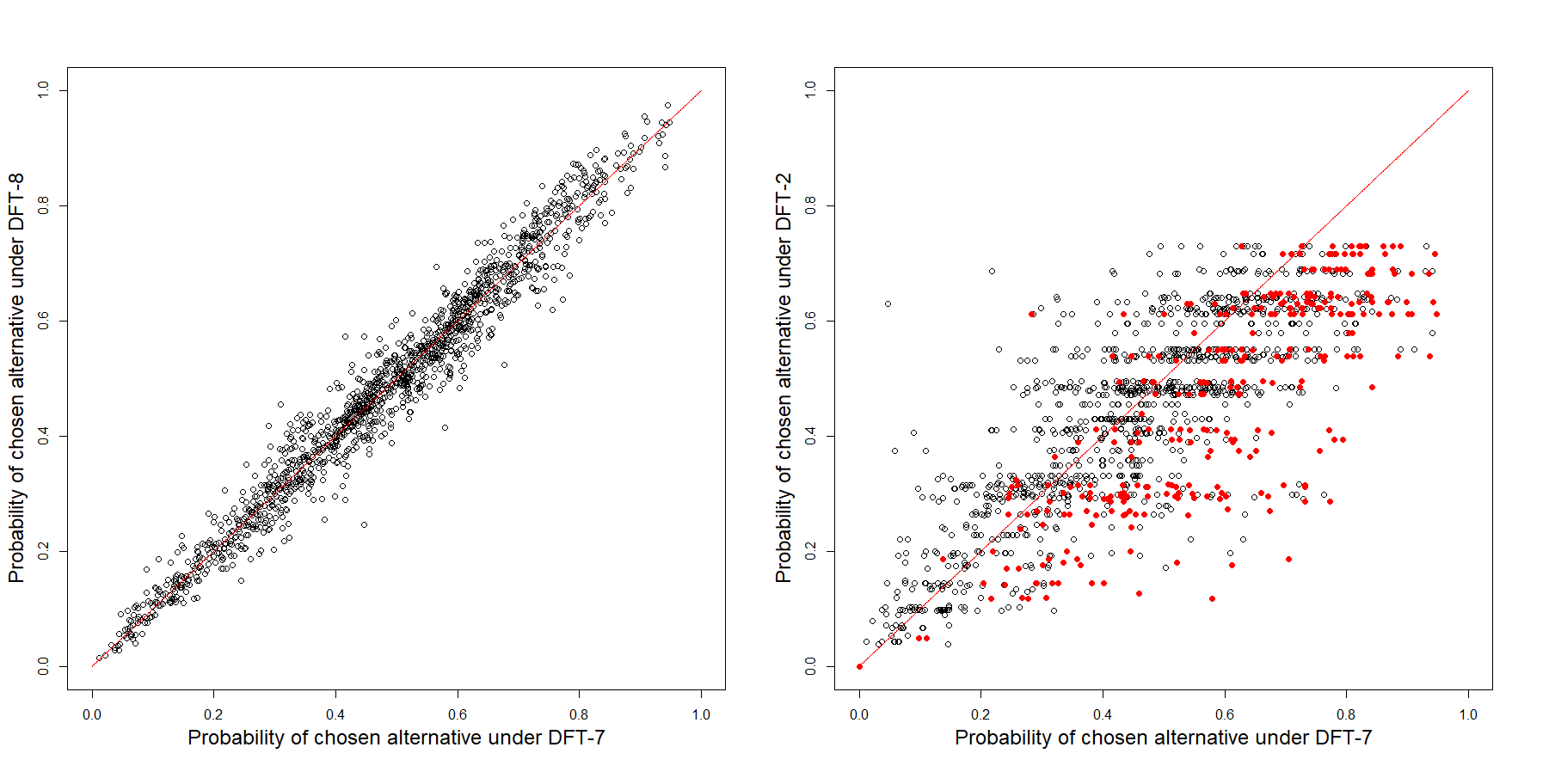}
            	\end{figure}

            \section{Empirical application: dynamic choice scenario (gap-acceptance)}
            
            For dynamic choice scenarios, the incorporation of additional information regarding choice deliberation is more complex. We consider data from a typical driving simulator experiment, where drivers choose whether to accept a gap or not when crossing an unsignalised intersection. First, we describe this data in detail. Next, we describe how we incorporate eye-tracking, heart rate and skin conductance responses (SCR) in both our MNL and DFT models. Finally, we analyse the results from adding heart rate and SCR data to our `basic' models without this information, and then also the addition of eye-tracking information. As before, the performance of the MNL and DFT based frameworks have been investigated.
            
                \subsection{Dataset}\label{sec:DynamicData}
                Our driving simulator dataset comes from \citet{paschalidis2018modelling}. 41 staff and student members of the University of Leeds completed a number of driving tasks in the University of Leeds driving simulator (UoLDS), a high fidelity, fully immersive, dynamic simulator (see Figure \ref{fig:Sim}). 
                
                \begin{figure}[ht!]
            		\centering
            		\caption{The University of Leeds driving simulator (sources: University of Leeds, University of Leeds driving simulator).}
            		\label{fig:Sim}
            		\includegraphics[scale=1]{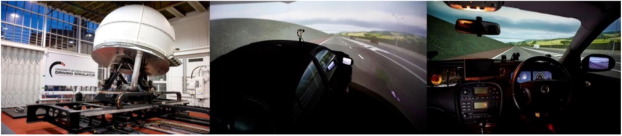}
            	\end{figure}
            	
            In this study, drivers negotiate a virtual world and face a number of scenarios designed to evoke stress, such as traffic lights that stay red and aggressive driving from other (computer controlled) vehicles. For the case in hand, we consider the two gap-acceptance tasks that each individual completes. For the first task, there is no time pressure, with drivers completing the tasks `naturally'. For the second, time pressure is induced to test whether drivers make different decisions and accept different gaps. The participants are told that they will receive less money for participating in the experiment if they finish `late'. Lateness is indicated by a red `unhappy' face that appears on the dashboard, though this face actually appears for the second set of driving tasks regardless of how quickly the first set of tasks are completed. The result is that when drivers are negotiating the second gap acceptance task, they believe that they are on course to miss out on some reward money. Full details of the simulator setup and details of the tasks are described by \citet{paschalidis2018modelling,paschalidis2019developing}. 
                
                To attempt to measure how stressed the drivers are when facing these different scenarios under different conditions, an Empatica E4 wristwatch is worn to collect heart rate (HR), electrodermal activity (EDA), blood volume pulse (BVP) and temperature (TMP) data. \citet{paschalidis2018modelling} considers these measures in detail and demonstrates that two variables are found to have a significant impact in the models. The first is heart rate data normalised at the individual level, with observations generated for the beginning of each gap, $y_{ni,hr}$, where the normalised heart rate (hr) for individual $n$ at gap $i$ is $y_{ni,hr} = (y_{ni,uhr} - y_{n,\mu hr})/y_{n,\sigma hr}$, where $y_{ni,uhr}$ corresponds to the observed heart rate value, $y_{n,\mu hr}$ is the participant's heart rate mean value across the set of tasks and $y_{n,\sigma hr}$ is the corresponding standard deviation. The second variable is based on a transformation of the EDA data using trough-to-peak analysis \citep{benedek2010continuous} to generate skin conductance responses (SCR), $y_{ni,scr}$, for each individual at each gap.
                
                Eye-tracking information was recorded using a v4.5 Seeing Machines FaceLab eye-tracker, recording at 60Hz. Gaze patterns have been explored in a number of studies using the UoLDS, including understanding a passenger's visual attention when in an autonomous vehicle \citep{louw2017you}, testing the impact of gaze concentration on lane keeping behaviour \citep{li2018improved} and understanding the sequence of transitions during lane-changes \citep{gonccalves2019using}. For the purposes of this study, there are a number of variables of interest, including how often the driver looks at oncoming vehicles when crossing the intersection (percentage of time looking to the left) and how distracted they are: whether they focus on one spot, i.e. a particular gap, or scan their full surroundings. As with the heart rate and SCR data, we require individual-level normalisations of gaze movement. 
                
                The eye movement data recorded includes yaw and pitch for each fixation, which determine whether a participant is looking to the left or right and up or down, respectively. Head yaw and pitch (as demonstrated in Figure \ref{fig:YawPitch}) as well as yaw and pitch for the gaze direction for both eyes are recorded.
                
                 \begin{figure}[ht!]
            		\centering
            		\caption{The angles of movement determining which direction the driver is looking (Figure source: \citealt{arcoverde2014enhanced}).}
            		\label{fig:YawPitch}
            		\includegraphics[scale=0.5]{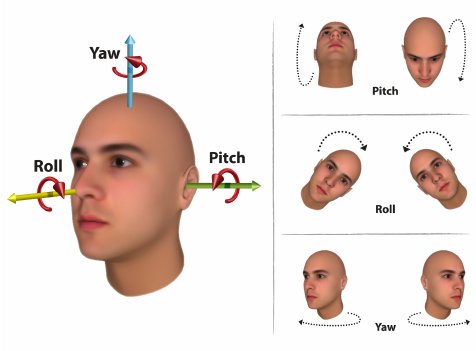}
            	\end{figure}
            	
                We thus use an aggregated measure to determine where driver $n$ is looking at time point $t$: 
            	\begin{linenomath*}
            		\begin{equation}
            			yaw_{nt} = yaw_{nt,head} + 0.5 \cdot yaw_{nt,left} + 0.5 \cdot yaw_{nt,right},
            		\end{equation}
            	\end{linenomath*}
                where $yaw_{nt,head}$ is the head yaw and  $yaw_{nt,left}$ and $yaw_{nt,right}$ are the yaw angles for the gaze direction of the left and right eyes respectively. $pitch_{nt}$ is calculated equivalently with head, left and right eye gaze pitches. 
                
                As we wish to normalise the yaws and pitches for each individual, the fixations are recalibrated such that both $yaw_{nt}$ and $pitch_{nt}$ have means of zero over the course of the two gap acceptance tasks. Each fixation is then assigned to one of four quadrants (left, right, up, down), with $pitch=yaw$ and $pitch= -yaw$ being used to define each quadrant. Fixations where $yaw_{nt}^2 + pitch_{nt}^2 < 36$ are defined to be in a fifth `centre' zone. An illustration of this is given in Figure \ref{fig:Gaze}, which gives the fixations after calibration for participant 1.
                
                \begin{figure}[ht!]
            		\centering
            		\caption{The gaze fixations during the gap acceptance tasks for participant 1.}
            		\label{fig:Gaze}
            		\includegraphics[scale=0.4]{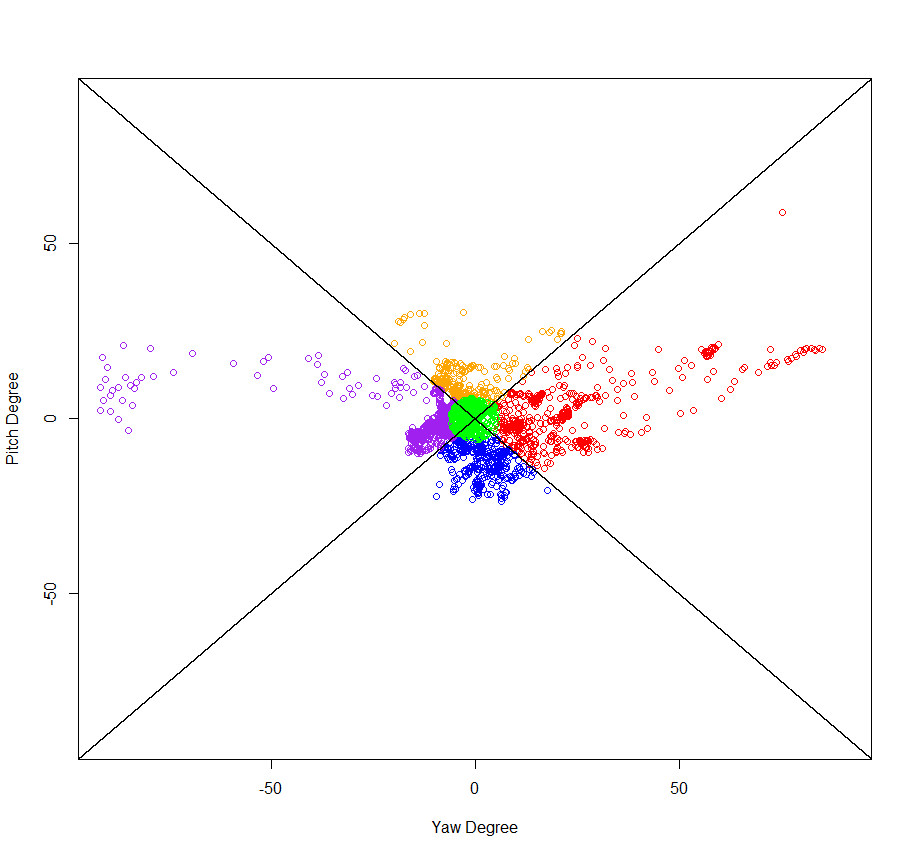}
            	\end{figure}
                
                For each gap that the driver faces, we can then define the percentage of time that a driver is looking in each zone, with $y_{n,gaze_{left,share}}$ for example representing the total proportion of time that the driver looks to the left. 
  
                \subsection{Base model}
                
                \subsubsection{Model specifications}
                After some initial specification testing, our first multinomial logit model (MNL) is based on three key attributes describing the gap acceptance task: the size of the gap between vehicles ($x_{n,gapsize}$), the position of the driver's vehicle in relation to the junction ($x_{n,pos}$), and the speed at which the driver arrives at the junction ($x_{n,speed}$). The utility for choosing to accept gap number $i$ for individual $n$ for is thus:
                
                \begin{linenomath*}
            	    \begin{subequations}\label{eq:MNL1}
			            \begin{align}
            		    U_{ni} &= \delta_G + (\beta_{gap} + \beta_{gapn} \cdot ln(i)) \cdot x_{n,gapsize} \\
            		    &+ \beta_{pos} \cdot x_{n,pos} \\
                        &+ \beta_{speed} \cdot (i=1) \cdot x_{n,speed} \\
            		    &+ \alpha_{age} \cdot z_{n,age} + \alpha_{reg} \cdot z_{n,reg} + \epsilon_{ni},
            		    \end{align}
            		\end{subequations}
            	\end{linenomath*}
                where $\delta_G$ is an alternative specific constant for accepting the gap, $\beta_{gap}$ gives the relative importance of the size of the gap and $\beta_{gapn}$ controls for the fact that the driver may over time become impatient and start considering accepting a smaller gap (where initial testing found that a logarithmic transform of the gap number improved model fit). $\beta_{speed}$ is the relative importance of the speed at which the driver arrives at the junction, and is only applied for the first gap. This allows the model to recognise that some drivers may `race' across as soon as possible. $\beta_{pos}$ gives the relative importance of the position of the vehicle, recognising that drivers may be more likely to accept a gap if they are closer to the crossing point. Finally, two sociodemographics are included through the use of dummy variables $z_{n,age}$ and $z_{n,reg}$, following the results of \citet{paschalidis2018modelling}. $\alpha_{age}$ is added if participant $n$ is over 45 and $\alpha_{reg}$ is added if the participant is a regular driver (defined as typically driving everyday).  
                
                Our DFT model is based on the same initial specification. Based on the findings of \citet{hancock2021accumulation}, we initially set up the model with estimated scaling parameters rather than estimated weights. We define three possible `attributes' that the driver considers: the size of the gap, the position of their vehicle and alternative factors. The attribute matrix for individual $n$ for gap $i$ is thus:
                
                \begin{linenomath*}
            	    \begin{equation}\label{eq:DFT1}
            		    M_{ni} = \begin{bmatrix}
            		    x_{gapsize} & x_{pos} & 1 \\
            		    0 & 0 & 0 
            		    \end{bmatrix},
            		\end{equation}
            	\end{linenomath*}
                
                with corresponding scaling factors for the gap ($s_{gap}$), the position ($s_{pos}$) and alternative factors ($s_{alt}$):
                
                \begin{linenomath*}
            	    \begin{equation}\label{eq:DFT2}
            		    \bm{\beta}  = \begin{bmatrix}
            		    s_{gap}\\
            		    s_{pos}\\
            		    s_{alt}
            		    \end{bmatrix} = \begin{bmatrix}
            		    \beta_{gap} + \beta_{gapn} \cdot ln(i) \\
            		    \beta_{pos} \\
            		    \delta_{bias} + \beta_{speed} \cdot (i=1) \cdot x_{n,speed} + \alpha_{age} \cdot z_{n,age} + \alpha_{reg} \cdot z_{n,reg}
            		    \end{bmatrix},
            		\end{equation}
            	\end{linenomath*}
                
                where the equivalent terms to the MNL model are used, with the addition of a $\delta_{bias}$ term. This parameter performs a similar function to an alternative specific constant in a random utility model. However, under DFT, an individual could have an initial preference towards an alternative ($\delta_{G}$) but then also have a bias towards an alternative that increases over the course of deliberating over the set of alternatives ($\delta_{bias}$) with these parameters being separately identifiable \citep{hancock2018decision}. DFT also estimates $\tau$, the parameter for the number of preference updating steps. In line with the guidance for fitting DFT models to choice tasks with only two alternatives (detailed in Table 1 in \citealt{hancock2021accumulation}), the error term $\sigma_\epsilon^2$ is fixed to a value of 1, the `attribute' attention weights $w_{gap}$, $w_{pos}$ and $w_{alt}$ are all fixed to a value of 1/3, the memory parameter $\phi_2$ is fixed to zero and the sensitivity parameter $\phi_1$ is not estimated. Thus, the DFT model incorporates all of the parameters included in the MNL model but has two extra estimated parameters, $\delta_{bias}$ and $\tau$.
                
                \subsubsection{Results}
                The results of our base MNL and DFT models are given in Table \ref{tab:BasicModels}.
                
                \begin{table}[ht!]
                  \centering
                  \scriptsize
                  \caption{Base model (without physiological data included) results for the gap acceptance data.}
                           \begin{tabular}{|c|cccc|}
                        \toprule
                        Model & \multicolumn{2}{c|}{MNL-B} & \multicolumn{2}{c|}{DFT-B} \\
                        \midrule
                        Description & \multicolumn{4}{c|}{Base models} \\
                        \midrule Log-likelihood(null) & \multicolumn{4}{c|}{-426.29} \\
                        \midrule    Log-likelihood & \multicolumn{2}{c|}{-117.98} & \multicolumn{2}{c|}{-101.27} \\
                            Estimated parameters & \multicolumn{2}{c|}{7} & \multicolumn{2}{c|}{9} \\
                            Adj. $\rho^2$ & \multicolumn{2}{c|}{0.7068} & \multicolumn{2}{c|}{0.7413} \\
                            BIC   & \multicolumn{2}{c|}{280.92} & \multicolumn{2}{c|}{260.33} \\
                            \midrule
                                  & \multicolumn{1}{c}{est.} & \multicolumn{1}{c|}{rob. t-rat.} & est.  & rob. t-rat. \\
                            \midrule
                            $\delta_G$ & \multicolumn{1}{c}{-11.44} & \multicolumn{1}{c|}{-4.58} & -229.12 & -6.21 \\
                            $\beta_{gap}$ & \multicolumn{1}{c}{1.35} & \multicolumn{1}{c|}{2.07} & 1.98  & 2.65 \\
                            $\beta_{gapn}$ & \multicolumn{1}{c}{-0.28} & \multicolumn{1}{c|}{-1.56} & -0.47 & -2.52 \\
                            $\beta_{speed}$ & \multicolumn{1}{c}{1.44} & \multicolumn{1}{c|}{2.72} & 0.77  & 3.29 \\
                            $\beta_{pos}$ & \multicolumn{1}{c}{0.85} & \multicolumn{1}{c|}{3.46} & 1.03  & 3.29 \\
                            $\alpha_{age}$ & \multicolumn{1}{c}{-1.58} & \multicolumn{1}{c|}{-3.19} & -1.14 & -3.38 \\
                            $\alpha_{reg}$ & \multicolumn{1}{c}{1.69} & \multicolumn{1}{c|}{2.95} & 1.18  & 3.00 \\
                            \midrule
                            $\delta_{bias}$ & \multicolumn{1}{c}{\cellcolor[rgb]{ .851,  .851,  .851}} & \multicolumn{1}{c}{\cellcolor[rgb]{ .851,  .851,  .851}} & 8.36  & 3.71 \\
                            $\phi_1$ & \multicolumn{1}{c}{\cellcolor[rgb]{ .851,  .851,  .851}} & \multicolumn{1}{c}{\cellcolor[rgb]{ .851,  .851,  .851}} & 0.00  & \textbf{fixed} \\
                            $\phi_2$ & \multicolumn{1}{c}{\cellcolor[rgb]{ .851,  .851,  .851}} & \multicolumn{1}{c}{\cellcolor[rgb]{ .851,  .851,  .851}} & 0.00  & \textbf{fixed} \\
                            $\sigma_{\epsilon}$ & \multicolumn{1}{c}{\cellcolor[rgb]{ .851,  .851,  .851}} & \multicolumn{1}{c}{\cellcolor[rgb]{ .851,  .851,  .851}} & 1.00  & \textbf{fixed} \\
                            $\tau$ & \multicolumn{1}{c}{\cellcolor[rgb]{ .851,  .851,  .851}} & \multicolumn{1}{c}{\cellcolor[rgb]{ .851,  .851,  .851}} & 15.87 & 3.51 \\
                            \bottomrule
                            \end{tabular}%

                  \label{tab:BasicModels}%
                \end{table}%

                Our DFT model substantially outperforms the MNL model, finding a gain of 16.7 log-likelihood units at the cost of two additional parameters. 
                Both models show results that are in line with our expectations. The larger the gap, the closer the car is to the crossing and the faster the participant is driving on approaching the junction, the more likely the participant is to accept the gap. Additionally, we observe that the size of the gap reduces in importance over time, indicating that the driver possibly becomes impatient and is more willing to consider smaller gaps the longer they have to wait. In line with the results of \citet{paschalidis2018modelling}, younger and more regular drivers have a lower threshold for accepting a gap. The DFT model suggests that drivers have a strong initial preference for rejecting a gap, but the probability of accepting the gap increases over the course of deliberation.
                
                \subsection{Models with stress indicator data}
                The second set of models is built upon the base set by adding in information regarding how stressed the drivers are. 
                
                \subsubsection{Model framework}
                For both our MNL and DFT models, this is achieved through the addition of three parameters, for which we define total stress ($\alpha_{ni,stress}$) of participant $n$ when considering gap $i$ as:
                
                \begin{linenomath*}
            	    \begin{equation}\label{eq:Stress1}
            		    \alpha_{ni,stress} =  \delta_{stress} \cdot x_{n,scen} + \alpha_{hr} \cdot z_{ni,hr} + \alpha_{scr} \cdot z_{ni,scr},
            		\end{equation}
            	\end{linenomath*}
                
                where $x_{n,scen}$ takes a value of one if driver $n$ is completing the second set of driving tasks (where there is time pressure) and zero otherwise. $z_{ni,hr}$ and $z_{ni,scr}$ are the driver's normalised heart rate and skin conductance response at the beginning of facing gap $i$ (as outlined in Section \ref{sec:DynamicData} and described in detail by \citealt{paschalidis2018modelling}). Finally, the relative importance of these variables are captured through $\delta_{stress}$, $\alpha_{hr}$ and $\alpha_{scr}$, which are estimated parameters.
                
                For the MNL model, this information is incorporated simply through an adjustment to $\delta_G$, with $\delta_G^\star = \delta_G + \alpha_{ni,stress}$. For DFT, we consider two possibilities. The first option is through an equivalent adjustment to $\delta_{G}$, which similarly allows for a change in the likelihood of accepting or rejecting a gap depending on how stressed the driver is (model DFT-S1 in the following section). The second option is to estimate a separate process error term for each individual depending on their level of stress (DFT-S2):
                
                \begin{linenomath*}
            	    \begin{equation}\label{eq:Stress2}
            		    \sigma_{ni,\epsilon} =  exp(\alpha_{ni,stress}).
            		\end{equation}
            	\end{linenomath*}
                
                The higher the error variance in a DFT model, the less deterministic the choices are, thus positive estimates for the stress parameters would in this case indicate that a driver makes less predictable choices when they are stressed.
                
                \subsubsection{Results}
                
                The results of the three models that additionally incorporate stress indicator data are given in Table \ref{tab:StressModels}.
                
                \begin{table}[ht!]
                  \centering
                  \scriptsize
                  \caption{Model results from incorporating stress level information for the gap acceptance data.}
                        \begin{tabular}{|c|cc|cc|cc|}
                            \toprule
                            Model & \multicolumn{2}{c|}{MNL-S} & \multicolumn{2}{c|}{DFT-S1} & \multicolumn{2}{c|}{DFT-S2} \\
                            \midrule
                            Base model & \multicolumn{2}{c|}{MNL-B} & \multicolumn{2}{c|}{DFT-B} & \multicolumn{2}{c|}{DFT-B} \\
                            \midrule
                            Additional information & \multicolumn{2}{c|}{stress added to U} & \multicolumn{2}{c|}{stress added to $P_0$} & \multicolumn{2}{c|}{$\sigma_{\epsilon}$ = f(stress)} \\
                            \midrule Log-likelihood(null) & \multicolumn{6}{c|}{-426.29} \\
                            \cmidrule{2-7}    Log-likelihood &         \multicolumn{2}{c|}{-115.00} &         \multicolumn{2}{c|}{-95.60} &         \multicolumn{2}{c|}{-90.93} \\
                            Estimated parameters & \multicolumn{2}{c|}{10} & \multicolumn{2}{c|}{12} & \multicolumn{2}{c|}{12} \\
                            Improvement over base model & \multicolumn{2}{c|}{2.98} & \multicolumn{2}{c|}{5.67} & \multicolumn{2}{c|}{10.34} \\
                            Likelihood ratio test p-value & \multicolumn{2}{c|}{0.11360} & \multicolumn{2}{c|}{0.01002} & \multicolumn{2}{c|}{0.00012} \\
                            Adj. $\rho^2$ & \multicolumn{2}{c|}{0.7068} & \multicolumn{2}{c|}{0.7476} & \multicolumn{2}{c|}{0.7585} \\
                            BIC   & \multicolumn{2}{c|}{294.22} & \multicolumn{2}{c|}{268.27} & \multicolumn{2}{c|}{258.91} \\
                            \midrule
                                  & est.  & rob. t-rat. & est.  & rob. t-rat. & est.  & rob. t-rat. \\
                            \midrule
                            $\delta_G$ & -11.70 & -4.60 & -267.28 & -4.18 & -544.25 & -4.10 \\
                            $\beta_{gap}$ & 1.30  & 1.90  & 2.15  & 3.80  & 3.53  & 2.30 \\
                            $\beta_{gapn}$ & -0.26 & -1.40 & -0.53 & -3.56 & -0.82 & -2.16 \\
                            $\beta_{speed}$ & 1.44  & 2.68  & 0.92  & 2.74  & 1.77  & 2.88 \\
                            $\beta_{pos}$ & 0.86  & 3.39  & 1.18  & 4.73  & 2.05  & 2.87 \\
                            $\alpha_{age}$ & -1.50 & -2.89 & -1.15 & -2.49 & -2.11 & -2.48 \\
                            $\alpha_{reg}$ & 1.69  & 3.00  & 1.25  & 2.70  & 2.25  & 2.93 \\
                            \midrule
                            $\delta_{bias}$ & \cellcolor[rgb]{ .851,  .851,  .851} & \cellcolor[rgb]{ .851,  .851,  .851} & 9.14  & 4.65  & 15.90 & 2.99 \\
                            $\phi_1$ & \cellcolor[rgb]{ .851,  .851,  .851} & \cellcolor[rgb]{ .851,  .851,  .851} & 0.00  & \textbf{fixed} & 0.00  & \textbf{fixed} \\
                            $\phi_2$ & \cellcolor[rgb]{ .851,  .851,  .851} & \cellcolor[rgb]{ .851,  .851,  .851} & 0.00  & \textbf{fixed} & 0.00  & \textbf{fixed} \\
                            $\sigma_{\epsilon}$ & \cellcolor[rgb]{ .851,  .851,  .851} & \cellcolor[rgb]{ .851,  .851,  .851} & 1.00  & \textbf{fixed} & \cellcolor[rgb]{ .851,  .851,  .851} & \cellcolor[rgb]{ .851,  .851,  .851} \\
                            $\tau$ & \cellcolor[rgb]{ .851,  .851,  .851} & \cellcolor[rgb]{ .851,  .851,  .851} & 16.78 & 4.47  & 19.80 & 2.83 \\
                            \midrule
                            $\delta_{stress}$ & 0.50  & 2.51  & 0.00  & -0.04 & 0.55  & 1.01 \\
                            $\alpha_{hr}$ & 0.12  & 0.52  & 0.25  & 1.35  & 0.91  & 4.37 \\
                            $\alpha_{scr}$ & 2.26  & 1.32  & 2.12  & 1.84  & 4.49  & 3.89 \\
                            \bottomrule
                        \end{tabular}%
                  \label{tab:StressModels}%
                \end{table}%
                
                Our MNL and first DFT specifications, which are based on the addition of stress measures to the underlying bias towards choosing whether to accept a gap or not, do not obtain large gains in log-likelihood. A likelihood ratio test of model MNL-S against MNL-B reveals that the new model can be rejected. For DFT-S1, a p-value of $0.01$ suggests some evidence in favour of the new model, though it has worse adjusted $\rho^2$ and BIC values than DFT-B. However, DFT-S2, which incorporates stress measures on the process error, records a significantly better model fit than DFT-B, with nearly twice the improvement in log-likelihood in comparison to DFT-S1. All three new parameters are positive, with $\alpha_{hr}$ and $\alpha_{scr}$ both being significant. This implies that within this model, drivers under measurable stress make less predictable decisions. The remaining parameters in DFT-S2 appear to have a scale difference in comparison to DFT-S1, with most parameters approximately doubled. This allows DFT-S2 to make more precise predictions than DFT-B for drivers who are not stressed. An illustration of this is given in Figure \ref{fig:Pressure}, where orange points represent a positive value for $\alpha_{ni,stress}$ under DFT-S2, indicating that the decision-maker is feeling stressed, whereas purple points represent situations in which the driver is not stressed. Circles represent `accept gap' decisions and crosses represent `reject gap' decisions. The black line indicates the point at which both models generate the same probability. The scale difference for DFT-S2 parameters results in a cluster of purple crosses above the line in the top right quadrant, suggesting that DFT-S2 does particularly well at predicting reject decisions when the driver has low stress.
                                
                \begin{figure}[ht!]
            		\centering
            		\caption{This figure compares the probabilities of the chosen alternative under models DFT-B and DFT-S2. }
            		\label{fig:Pressure}
            		\includegraphics[scale=0.5]{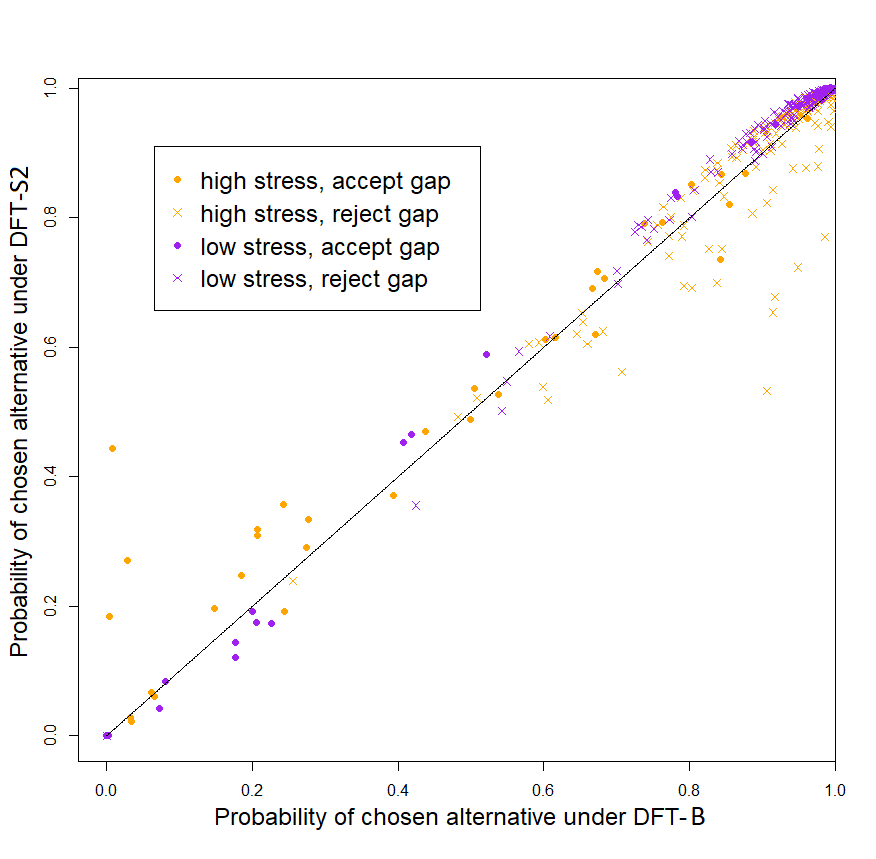}
            	\end{figure}
                                
                \subsection{Models incorporating eye-tracking information}
                
                Our final set of models additionally incorporate eye-tracking information. 
                
                \subsubsection{Model framework}
                
                We use three measures of gaze movement for each gap, with the first being based on the proportion of time the driver looks to the left, $y_{ni,gaze_{left}}$, which is defined:
                
                \begin{linenomath*}
            	    \begin{equation}\label{eq:Gaze1}
            		    y_{ni,gaze_{left}} =  y_{ni,gaze_{left-5}} - y_{n,gaze_{left-share}},
            		\end{equation}
            	\end{linenomath*}
                
                where $y_{ni,gaze_{left-5}}$ is the proportion of time the driver looks to the left in the 5 seconds before they first face gap $i$, and  $y_{n,gaze_{left-share}}$ is the proportion of time the driver looks to the left over the course of both gap acceptance scenarios. 
                
                The other two additional measures are based on how much the driver `explores' their environment. $y_{ni,gaze_{yaw-sd}}$ is the standard deviation of gaze yaw (how much the driver looks from left to right) and $y_{ni,gaze_{pitch-sd}}$ is the standard deviation of gaze pitch (how much the driver looks up and down), with both being measured across all fixations in the 5 seconds prior to the gap appearing. 
                
                These three measures are interacted with their respective estimated parameters ($\alpha_{gaze_{left}}$, $\alpha_{gaze_{yaw-sd}}$ and $\alpha_{gaze_{pitch-sd}}$) and added to give a total measure $\alpha_{ni,gaze}$:
                
                \begin{linenomath*}
            	    \begin{equation}\label{eq:Gaze2}
            		    \alpha_{ni,gaze}  = y_{ni,gaze_{left}} \cdot \alpha_{gaze_{left}} + y_{ni,gaze_{yaw-sd}} \cdot \alpha_{gaze_{yaw-sd}} + y_{ni,gaze_{pitch-sd}} \cdot \alpha_{gaze_{pitch-sd}}.
            		\end{equation}
            	\end{linenomath*}
                
                For the MNL model, this information is simply added to the total utility for accepting the gap, such that $\delta_G^{\star\star} = \delta_G^{\star} + \alpha_{ni,stress} + \alpha_{ni,gaze}$. For the DFT model, we consider three different options for incorporating the eye-tracking information. The first (DFT-E1), in line with the MNL specification, simply adds $\alpha_{ni,gaze}$ to $\delta_G$. The next two options consider whether it is more suitable to include eye-tracking information on DFT's scaling or weight parameters. For the 2nd option (DFT-E2), we adjust the first entry of $\bm{\beta}$ in Eq. \ref{eq:DFT2} to $\beta_{gap} + \beta_{gapn} \cdot ln(i) + \alpha_{ni,gaze}$. For the third (DFT-E3), we instead add to the attention weight for the gap size ($w_{gap}$, which is still not an estimated parameter), such that we have an attention vector, $\bm{w}_{ni}$, defined: 
                
               \begin{linenomath*}
            	    \begin{equation}\label{eq:Gaze3}
            		    \bm{w}_{ni}  = \begin{bmatrix}
            		    w_{ni,gap}\\
            		    w_{ni,pos}\\
            		    w_{ni,alt}
            		    \end{bmatrix} = \begin{bmatrix}
            		    exp(\alpha_{ni,gaze})/(2+exp(\alpha_{ni,gaze}) \\
            		    1/(2+exp(\alpha_{ni,gaze}) \\
            		    1/(2+exp(\alpha_{ni,gaze})
            		    \end{bmatrix} 
            		    ,
            		\end{equation}
            	\end{linenomath*}
            	 with the logarithmic functions used such that the three weights still sum to one for all choice contexts regardless of the value of $\alpha_{ni,gaze}$.

                \subsubsection{Results}
                
                The results of our models incorporating eye-tracking information are given in Table \ref{tab:EyeModels}.
                
                \begin{table}[ht!]
                  \centering
                  \scriptsize
                  \caption{Model results from incorporating eye-tracking information for the gap acceptance data}
                    \begin{tabular}{|c|cc|cc|cc|cc|}
                        \toprule
                        Model & \multicolumn{2}{c|}{MNL-E} & \multicolumn{2}{c|}{DFT-E1} & \multicolumn{2}{c|}{DFT-E2} & \multicolumn{2}{c|}{DFT-E3} \\
                        \midrule
                        Base Model & \multicolumn{2}{c|}{MNL-B} & \multicolumn{2}{c|}{DFT-S2} & \multicolumn{2}{c|}{DFT-S2} & \multicolumn{2}{c|}{DFT-S2} \\
                        \midrule
                        Additional information & \multicolumn{2}{c|}{$\alpha_{gaze}$ added to U} & \multicolumn{2}{c|}{$\alpha_{gaze}$ added to $P_0$} & \multicolumn{2}{c|}{$\alpha_{gaze}$ added to $\beta_{gap}$} & \multicolumn{2}{c|}{$\alpha_{gaze}$ added to $w_{gap}$} \\
                        \midrule
                        Log-likelihood(null) & \multicolumn{8}{c|}{-426.29} \\

                            \cmidrule{2-9}    Log-likelihood &     \multicolumn{2}{c|}{-104.69} &     \multicolumn{2}{c|}{-88.15} &     \multicolumn{2}{c|}{-88.85} &     \multicolumn{2}{c|}{-84.85} \\
                            Estimated parameters & \multicolumn{2}{c|}{13} & \multicolumn{2}{c|}{15} & \multicolumn{2}{c|}{15} & \multicolumn{2}{c|}{15} \\
                            Improvement over MNL-S/DFT-S2 & \multicolumn{2}{c|}{10.31} & \multicolumn{2}{c|}{2.78} & \multicolumn{2}{c|}{2.08} & \multicolumn{2}{c|}{6.08} \\
                            Likelihood ratio test p-value & \multicolumn{2}{c|}{0.00013} & \multicolumn{2}{c|}{0.13510} & \multicolumn{2}{c|}{0.24470} & \multicolumn{2}{c|}{0.00686} \\
                            Adj. $\rho^2$ & \multicolumn{2}{c|}{0.7239} & \multicolumn{2}{c|}{0.7580} & \multicolumn{2}{c|}{0.7564} & \multicolumn{2}{c|}{0.7658} \\
                            BIC   & \multicolumn{2}{c|}{292.85} & \multicolumn{2}{c|}{272.62} & \multicolumn{2}{c|}{274.03} & \multicolumn{2}{c|}{266.03} \\
                            \midrule
                                  & est.  & rob. t-rat. & est.  & rob. t-rat. & est.  & rob. t-rat. & est.  & rob. t-rat. \\
                            \midrule
                            $\delta_G$ & -13.59 & -4.67 & -616.61 & -3.25 & -691.74 & -2.43 & -668.20 & -2.29 \\
                            $\beta_{gap}$ & 1.61  & 2.03  & 4.15  & 1.84  & 4.44  & 2.24  & 7.08  & 1.74 \\
                            $\beta_{gapn}$ & -0.35 & -1.61 & -0.99 & -1.77 & -1.05 & -2.05 & -1.71 & -1.75 \\
                            $\beta_{speed}$ & 1.84  & 4.68  & 2.04  & 2.57  & 2.20  & 2.48  & 4.45  & 1.53 \\
                            $\beta_{pos}$ & 1.03  & 4.49  & 2.21  & 2.23  & 2.45  & 3.01  & 3.46  & 1.69 \\
                            $\alpha_{age}$ & -1.56 & -2.61 & -2.19 & -2.08 & -2.45 & -2.49 & -3.02 & -1.55 \\
                            $\alpha_{reg}$ & 1.77  & 3.01  & 2.36  & 2.29  & 2.65  & 2.42  & 3.24  & 1.92 \\
                            \midrule
                            $\delta_{bias}$ & \cellcolor[rgb]{ .851,  .851,  .851} & \cellcolor[rgb]{ .851,  .851,  .851} & 17.61 & 2.28  & 19.42 & 2.71  & 26.09 & 1.81 \\
                            $\phi_1$ & \cellcolor[rgb]{ .851,  .851,  .851} & \cellcolor[rgb]{ .851,  .851,  .851} & 0.00  & \textbf{fixed} & 0.00  & \textbf{fixed} & 0.00  & \textbf{fixed} \\
                            $\phi_2$ & \cellcolor[rgb]{ .851,  .851,  .851} & \cellcolor[rgb]{ .851,  .851,  .851} & 0.00  & \textbf{fixed} & 0.00  & \textbf{fixed} & 0.00  & \textbf{fixed} \\
                            $\sigma_{\epsilon}$ & \cellcolor[rgb]{ .851,  .851,  .851} & \cellcolor[rgb]{ .851,  .851,  .851} &\cellcolor[rgb]{ .851,  .851,  .851}   & \cellcolor[rgb]{ .851,  .851,  .851}  & \cellcolor[rgb]{ .851,  .851,  .851}   & \cellcolor[rgb]{ .851,  .851,  .851}  & \cellcolor[rgb]{ .851,  .851,  .851}   & \cellcolor[rgb]{ .851,  .851,  .851}  \\
                            $\tau$ & \cellcolor[rgb]{ .851,  .851,  .851} & \cellcolor[rgb]{ .851,  .851,  .851} & 20.28 & 2.71  & 20.70 & 3.12  & 14.08 & 2.87 \\
                            \midrule
                            $\delta_{stress}$ & 0.42  & 1.85  & 0.46  & 0.61  & 0.46  & 0.52  & 0.89  & 1.84 \\
                            $\alpha_{hr}$ & 0.06  & 0.25  & 0.87  & 3.17  & 0.95  & 2.98  & 0.81  & 2.93 \\
                            $\alpha_{scr}$ & 2.62  & 1.42  & 4.87  & 3.64  & 4.99  & 3.63  & 5.68  & 3.80 \\
                            \midrule
                            $\alpha_{gaze_{left}}$ & -10.76 & -3.36 & -122.27 & -2.14 & -2.12 & -2.19 & 61.05 & 2.89 \\
                            $\alpha_{gaze_{yaw-sd}}$ & 0.24  & 3.38  & 2.80  & 1.74  & 0.05  & 1.72  & -1.41 & -2.94 \\
                            $\alpha_{gaze_{pitch-sd}}$ & 0.20  & 1.29  & 4.34  & 1.38  & 0.04  & 0.90  & -2.44 & -4.16 \\
                            \bottomrule
                        \end{tabular}%
                  \label{tab:EyeModels}%
                \end{table}%

        For MNL-E and DFT-E1, we observe significant negative estimates for $\alpha_{gaze_{left}}$, indicating that the longer the driver spends looking left (relative to their average time spent looking left), the less likely the driver is to accept the gap. This could be a result of the driver looking at not only the upcoming gap, but future gaps, which they may spend more time on if they have already rejected or are planning on rejecting the current gap. The MNL model finds a significant improvement over the MNL model without eye-tracking information, whereas the first DFT model is not significantly better than the previous DFT model. The same can be said of DFT-E2, which incorporates eye-tracking information on the scaling parameters. As a contrast to the results for our static choice context, it is our model with eye-tracking information on the attention weight parameters results in a better fit in the dynamic case. DFT-E3, whilst recording a worse BIC, obtains a better adjusted $\rho^2$ value and a significantly better log-likelihood than DFT-S2. Notably, the parameter estimates for this model also give us insights into the decision-making process. Firstly, it finds a significant positive value for $\alpha_{gaze_{left}}$, implying that the longer the driver spends looking to the left, the greater the importance of the size of the gap. This intuitively makes sense - the driver spends more time looking at the gap, and hence its size becomes of greater importance in deciding whether the gap will be accepted or not. Secondly, we observe significant negative estimates for the two other gaze parameters, suggesting that the larger the standard deviations of gaze pitch and gaze yaw, the less important the size of the gap is. This is also psychologically plausible, as it implies that if the driver's attention is more focused (i.e. if they have smaller standard deviations, and are thus looking around less), the attention weight for the size of the gap increases. 
        
        \subsection{Summary of findings}
        In conclusion, these results show that DFT’s process noise parameter can be linked to stress indicators for modelling gap acceptance tasks, with model DFT-S2 in Table \ref{tab:StressModels} showing a significantly better performance than either DFT-B without stress indicators or DFT-S1 with stress indicators incorporated into initial preferences. Intuitively, it appears that more stressed decision-makers make less consistent choices, with higher noise estimated for decision-makers who are stressed. Furthermore, model DFT-E3 gives some interesting insights with regard to gaze patterns when a decision-maker is choosing whether to accept a gap or not. Again, intuitively, the results make sense, with a driver who looks more to the left (oncoming vehicles are coming from the left) having a higher estimated value for the attention weight for the size of the gap. Similarly, drivers who are more focused, with smaller standard deviations of gaze pitch and gaze yaw, assign a higher importance weight to the gap size. 
        
        It may be noted, though, that in this study, even though the models are dynamic, the process data is incorporated in a static format, using aggregated measures such as total time spent looking in one direction or at one attribute, rather than momentary fixations being used over the course of the deliberation process. Further steps towards a more dynamic model are possible. For example, a DFT model could incorporate a real `forgetting' function through simulating the updating preference process based on the attribute that is actually being looked at during each fixation. Similarly, a driver who looks to the left the second before they choose to accept a gap or not may make a different choice from a driver who looks a few seconds before needing to make the decision. The promising findings of the current study also serves as a motivation to integrate eye-tracking and skin conductance data in other driving behaviour models like lane-change and car-following. It may also provide insights for pedestrian and cyclist behaviour. Further investigation across a range of other contexts will help researchers to better understand the methods of incorporating physiological data that will consistently yield better choice models in general.
            
    \section{Conclusions}
        
        Whilst several studies have already demonstrated the use of eye-tracking data in choice models \citep{uggeldahl2016choice,krucien2017visual,van2018using}, no previous application has tested its use within psychological choice models that can simultaneously capture preferences across a large number of attributes. We thus provide a first comparison of whether incorporating this data has a different impact in psychological choice models in comparison with econometric choice models. We hypothesised that psychological choice models, which have `process’ parameters to conceptually capture the choice deliberation process, would be a better fit for the incorporation of `process’ data such as physiological data, due to the possibility that the process data can be linked to the process parameters. We explored this theory through a number of methods for the incorporation of physiological data within a DFT model for both static and dynamic choice contexts. We found that DFT records a better improvement in model fit in comparison to MNL, and that some of the physiological indicators can indeed be linked to process parameters. The results for the static scenario indicated that the addition of eye-tracking information to scaling parameters rather than attention weights results in better model performance. In contrast, the results for the dynamic scenario indicated that gaze pitch and yaw affect the attention weights, while the process noise parameters were better captured by the stress indicator data instead. This is not altogether surprising, given that these are very different choice contexts. The empirical findings thus indicate that the optimal framework for incorporating eye tracking data in choice models can vary depending on the application scenario and/or the type of choice that is investigated. Further investigation with a wider range of static and dynamic contexts would confirm if it is possible to develop a generic framework for incorporating eye-tracking data in choice models.
        
        However, it is clear from our results that the incorporation of physiological information into both psychological and econometric choice models offers fertile ground for future research in choice modelling and beyond. In particular, with the increased automation of cars, eye-tracking data for monitoring driver fatigue \citep{xu2018real} is expected to become available on a commercial basis in the near future, emphasising the importance of establishing best practices for the incorporation of such data in our choice models. 
        
        Directions for future research could include the use of eye-tracking data to simulate the precise updating of preferences (i.e. each updating step in Figure \ref{fig:DFTpref} could be represented by a fixation) to further test whether the dynamic nature of models such as DFT (or other models that incorporate an information search process, such as `RUM-DFT', \citealp{nova2025random}) help explain choice behaviour. Further work  
        should also include investigation of alternative frameworks, specifically including the incorporation of eye-tracking data as latent variables \citep{bansal2024discrete} to enhance psychological choice models as well as investigating the usefulness of such data in comparison and/or conjunction with brain imaging data such as EEG.        

        \section*{Acknowledgements}
	Thomas Hancock and Stephane Hess acknowledge the ﬁnancial support by the European Research Council through the advanced grant 101020940-SYNERGY.    Thomas Hancock and Charisma Choudhury’s time were partially supported by UKRI Future Leader Fellowship [MRT020423/1].

	    \bibliography{DFT_eye}

\begin{thebibliography}{}

\bibitem[Arcoverde~Neto et~al., 2014]{arcoverde2014enhanced}
Arcoverde~Neto, E.~N., Duarte, R.~M., Barreto, R.~M., Magalh{\~a}es, J.~P.,
  Bastos, C., Ren, T.~I., and Cavalcanti, G.~D. (2014).
\newblock Enhanced real-time head pose estimation system for mobile device.
\newblock {\em Integrated Computer-Aided Engineering}, 21(3):281--293.

\bibitem[Bansal et~al., 2024]{bansal2024discrete}
Bansal, P., Kim, E.-J., and Ozdemir, S. (2024).
\newblock Discrete choice experiments with eye-tracking: How far we have come
  and ways forward.
\newblock {\em Journal of choice modelling}, 51:100478.

\bibitem[Barr{\'\i}a et~al., 2023]{barria2023relating}
Barr{\'\i}a, C., Guevara, C.~A., Jimenez-Molina, A., and Seriani, S. (2023).
\newblock Relating emotions, psychophysiological indicators and context in
  public transport trips: Case study and a joint framework for data collection
  and analysis.
\newblock {\em Transportation Research Part F: Traffic Psychology and
  Behaviour}, 95:418--431.

\bibitem[Ben-Akiva et~al., 2002]{ben2002hybrid}
Ben-Akiva, M., McFadden, D., Train, K., Walker, J., Bhat, C., Bierlaire, M.,
  Bolduc, D., Boersch-Supan, A., Brownstone, D., Bunch, D.~S., et~al. (2002).
\newblock Hybrid choice models: progress and challenges.
\newblock {\em Marketing Letters}, 13(3):163--175.

\bibitem[Benedek and Kaernbach, 2010]{benedek2010continuous}
Benedek, M. and Kaernbach, C. (2010).
\newblock A continuous measure of phasic electrodermal activity.
\newblock {\em Journal of neuroscience methods}, 190(1):80--91.

\bibitem[Berkowitsch et~al., 2014]{berkowitsch2014rigorously}
Berkowitsch, N.~A., Scheibehenne, B., and Rieskamp, J. (2014).
\newblock Rigorously testing multialternative decision field theory against
  random utility models.
\newblock {\em Journal of Experimental Psychology: General}, 143(3):1331.

\bibitem[Bogacz et~al., 2019]{bogacz2019modelling}
Bogacz, M., Hess, S., Choudhury, C., Calastri, C., Erath, A., Van~Eggermond,
  M., Mushtaq, F., and Awais, M. (2019).
\newblock Modelling risk perception using a dynamic hybrid choice model and
  brain-imaging data: application to virtual reality cycling.
\newblock In {\em International Choice Modelling Conference 2019}.

\bibitem[Brown and Heathcote, 2008]{brown2008simplest}
Brown, S.~D. and Heathcote, A. (2008).
\newblock The simplest complete model of choice response time: Linear ballistic
  accumulation.
\newblock {\em Cognitive Psychology}, 57(3):153--178.

\bibitem[Busemeyer et~al., 2019]{busemeyer2019cognitive}
Busemeyer, J.~R., Gluth, S., Rieskamp, J., and Turner, B.~M. (2019).
\newblock Cognitive and neural bases of multi-attribute, multi-alternative,
  value-based decisions.
\newblock {\em Trends in cognitive sciences}, 23(3):251--263.

\bibitem[Busemeyer and Rieskamp, 2014]{busemeyer2014psychological}
Busemeyer, J.~R. and Rieskamp, J. (2014).
\newblock Psychological research and theories on preferential choice.
\newblock In {\em Handbook of choice modelling}. Edward Elgar Publishing.

\bibitem[Busemeyer and Townsend, 1992]{busemeyer1992fundamental}
Busemeyer, J.~R. and Townsend, J.~T. (1992).
\newblock Fundamental derivations from decision field theory.
\newblock {\em Mathematical Social Sciences}, 23(3):255--282.

\bibitem[Busemeyer and Townsend, 1993]{busemeyer1993decision}
Busemeyer, J.~R. and Townsend, J.~T. (1993).
\newblock Decision field theory: a dynamic-cognitive approach to decision
  making in an uncertain environment.
\newblock {\em Psychological Review}, 100(3):432.

\bibitem[Cohen et~al., 2017]{cohen2017multi}
Cohen, A.~L., Kang, N., and Leise, T.~L. (2017).
\newblock Multi-attribute, multi-alternative models of choice: Choice, reaction
  time, and process tracing.
\newblock {\em Cognitive Psychology}, 98:45--72.

\bibitem[Diederich, 1997]{diederich1997dynamic}
Diederich, A. (1997).
\newblock Dynamic stochastic models for decision making under time constraints.
\newblock {\em Journal of Mathematical Psychology}, 41(3):260--274.

\bibitem[Fisher, 2017]{fisher2017attentional}
Fisher, G. (2017).
\newblock An attentional drift diffusion model over binary-attribute choice.
\newblock {\em Cognition}, 168:34--45.

\bibitem[Foss and Goodwin, 2014]{foss2014distracted}
Foss, R.~D. and Goodwin, A.~H. (2014).
\newblock Distracted driver behaviors and distracting conditions among
  adolescent drivers: Findings from a naturalistic driving study.
\newblock {\em Journal of Adolescent Health}, 54(5):S50--S60.

\bibitem[Gigerenzer and Gaissmaier, 2011]{gigerenzer2011heuristic}
Gigerenzer, G. and Gaissmaier, W. (2011).
\newblock Heuristic decision making.
\newblock {\em Annual review of psychology}, 62:451--482.

\bibitem[Gl{\"o}ckner et~al., 2012]{glockner2012processing}
Gl{\"o}ckner, A., Fiedler, S., Hochman, G., Ayal, S., and Hilbig, B. (2012).
\newblock Processing differences between descriptions and experience: A
  comparative analysis using eye-tracking and physiological measures.
\newblock {\em Frontiers in psychology}, 3:173.

\bibitem[Gon{\c{c}}alves et~al., 2019]{gonccalves2019using}
Gon{\c{c}}alves, R., Louw, T., Madigan, R., and Merat, N. (2019).
\newblock Using markov chains to understand the sequence of drivers' gaze
  transitions during lane-changes in automated driving.
\newblock In {\em Proceedings of the... international driving symposium on
  human factors in driver assessment, training and vehicle design}, volume
  2019, pages 217--223. University of Iowa Public Policy Center.

\bibitem[Gonz{\'a}lez-Vald{\'e}s and Ortuzar, 2018]{gonzalez2018stochastic}
Gonz{\'a}lez-Vald{\'e}s, F. and Ortuzar, J. {\relax de}.~D. (2018).
\newblock The stochastic satisficing model: A bounded rationality discrete
  choice model.
\newblock {\em Journal of choice modelling}, 27:74--87.

\bibitem[Hancock and Choudhury, 2023]{hancock2023utilising}
Hancock, T.~O. and Choudhury, C.~F. (2023).
\newblock Utilising physiological data for augmenting travel choice models:
  methodological frameworks and directions of future research.
\newblock {\em Transport reviews}, 43(5):838--866.

\bibitem[Hancock et~al., 2018]{hancock2018decision}
Hancock, T.~O., Hess, S., and Choudhury, C.~F. (2018).
\newblock Decision field theory: Improvements to current methodology and
  comparisons with standard choice modelling techniques.
\newblock {\em Transportation Research Part B: Methodological}, 107:18--40.

\bibitem[Hancock et~al., 2021]{hancock2021accumulation}
Hancock, T.~O., Hess, S., Marley, A., and Choudhury, C.~F. (2021).
\newblock An accumulation of preference: Two alternative dynamic models for
  understanding transport choices.
\newblock {\em Transportation Research Part B: Methodological}, 149:250--282.

\bibitem[Henriquez-Jara and Guevara, 2025]{henriquez2025experience}
Henriquez-Jara, B. and Guevara, C.~A. (2025).
\newblock An experience-based choice model (ebcm): Formulation, identification,
  behavioural insights and well-being assessment.
\newblock {\em Journal of Choice Modelling}, 55:100552.

\bibitem[Henriquez-Jara et~al., 2025]{henriquez2025identifying}
Henriquez-Jara, B., Guevara, C.~A., and Jimenez-Molina, A. (2025).
\newblock Identifying instant utility using psychophysiological indicators in a
  transport experiment with ecological validity.
\newblock {\em Transportation}, pages 1--25.

\bibitem[Henr{\'\i}quez-Jara et~al., 2025]{henriquez4988311modelling}
Henr{\'\i}quez-Jara, B., Hancock, T.~O., Solernou, A., Garc{\'\i}a, J.,
  Guevara, A., and Choudhury, C.~F. (2025).
\newblock Modelling the effect of travel experiences in modal choice using
  virtual reality and physiological sensor data.
\newblock {\em Transporation Research Part C (forthcoming)}.

\bibitem[Hess and Stathopoulos, 2013]{hess2013linking}
Hess, S. and Stathopoulos, A. (2013).
\newblock Linking response quality to survey engagement: a combined random
  scale and latent variable approach.
\newblock {\em Journal of Choice Modelling}, 7:1--12.

\bibitem[Huang et~al., 2015]{huang2015age}
Huang, Y.~H., Wood, S., Berger, D.~E., and Hanoch, Y. (2015).
\newblock Age differences in experiential and deliberative processes in
  unambiguous and ambiguous decision making.
\newblock {\em Psychology and aging}, 30(3):675.

\bibitem[Hurts et~al., 2011]{hurts2011distracted}
Hurts, K., Angell, L.~S., and Perez, M.~A. (2011).
\newblock The distracted driver: Mechanisms, models, and measurement.
\newblock {\em Reviews of human factors and ergonomics}, 7(1):3--57.

\bibitem[Krajbich et~al., 2012]{krajbich2012attentional}
Krajbich, I., Lu, D., Camerer, C., and Rangel, A. (2012).
\newblock The attentional drift-diffusion model extends to simple purchasing
  decisions.
\newblock {\em Frontiers in psychology}, 3:193.

\bibitem[Krucien et~al., 2017]{krucien2017visual}
Krucien, N., Ryan, M., and Hermens, F. (2017).
\newblock Visual attention in multi-attributes choices: What can eye-tracking
  tell us?
\newblock {\em Journal of Economic Behavior \& Organization}, 135:251--267.

\bibitem[Li et~al., 2018a]{li2018improved}
Li, P., Markkula, G., Li, Y., and Merat, N. (2018a).
\newblock Is improved lane keeping during cognitive load caused by increased
  physical arousal or gaze concentration toward the road center?
\newblock {\em Accident Analysis \& Prevention}, 117:65--74.

\bibitem[Li et~al., 2018b]{li2018does}
Li, P., Merat, N., Zheng, Z., Markkula, G., Li, Y., and Wang, Y. (2018b).
\newblock Does cognitive distraction improve or degrade lane keeping
  performance? analysis of time-to-line crossing safety margins.
\newblock {\em Transportation research part F: traffic psychology and
  behaviour}, 57:48--58.

\bibitem[Litvak et~al., 2010]{litvak2010fuel}
Litvak, P.~M., Lerner, J.~S., Tiedens, L.~Z., and Shonk, K. (2010).
\newblock Fuel in the fire: How anger impacts judgment and decision-making.
\newblock In {\em International handbook of anger}, pages 287--310. Springer.

\bibitem[Louw and Merat, 2017]{louw2017you}
Louw, T. and Merat, N. (2017).
\newblock Are you in the loop? using gaze dispersion to understand driver
  visual attention during vehicle automation.
\newblock {\em Transportation Research Part C: Emerging Technologies},
  76:35--50.

\bibitem[Maule et~al., 2000]{maule2000effects}
Maule, A.~J., Hockey, G. R.~J., and Bdzola, L. (2000).
\newblock Effects of time-pressure on decision-making under uncertainty:
  changes in affective state and information processing strategy.
\newblock {\em Acta psychologica}, 104(3):283--301.

\bibitem[McFadden, 1974]{mcfadden1974conditional}
McFadden, D. (1974).
\newblock {Conditional Logit Analysis of Qualitative Choice Behaviour. In
  Frontiers in Econometrics, ed. P. Zarembka} ({N}ew {Y}ork: Academic press).

\bibitem[Noguchi and Stewart, 2018]{noguchi2018multialternative}
Noguchi, T. and Stewart, N. (2018).
\newblock Multialternative decision by sampling: A model of decision making
  constrained by process data.
\newblock {\em Psychological review}, 125(4):512.

\bibitem[Nova et~al., 2025]{nova2025random}
Nova, G., Guevara, A., Hess, S., and Hancock, T.~O. (2025).
\newblock Random utility maximisation model considering the information search
  process.
\newblock {\em Transportation Research Part B (forthcoming)}.

\bibitem[Nunez et~al., 2017]{nunez2017attention}
Nunez, M.~D., Vandekerckhove, J., and Srinivasan, R. (2017).
\newblock How attention influences perceptual decision making: Single-trial eeg
  correlates of drift-diffusion model parameters.
\newblock {\em Journal of mathematical psychology}, 76:117--130.

\bibitem[Paschalidis, 2019]{paschalidis2019developing}
Paschalidis, E. (2019).
\newblock {\em Developing driving behaviour models incorporating the effects of
  stress}.
\newblock PhD thesis, University of Leeds.

\bibitem[Paschalidis et~al., 2018]{paschalidis2018modelling}
Paschalidis, E., Choudhury, C.~F., and Hess, S. (2018).
\newblock Modelling the effects of stress on gap-acceptance decisions combining
  data from driving simulator and physiological sensors.
\newblock {\em Transportation research part F: traffic psychology and
  behaviour}, 59:418--435.

\bibitem[Paschalidis et~al., 2019]{paschalidis2019combining}
Paschalidis, E., Choudhury, C.~F., and Hess, S. (2019).
\newblock Combining driving simulator and physiological sensor data in a latent
  variable model to incorporate the effect of stress in car-following
  behaviour.
\newblock {\em Analytic methods in accident research}, 22:100089.

\bibitem[Peters et~al., 2007]{peters2007adult}
Peters, E., Hess, T.~M., V{\"a}stfj{\"a}ll, D., and Auman, C. (2007).
\newblock Adult age differences in dual information processes: Implications for
  the role of affective and deliberative processes in older adults' decision
  making.
\newblock {\em Perspectives on Psychological Science}, 2(1):1--23.

\bibitem[Pettibone, 2012]{pettibone2012testing}
Pettibone, J.~C. (2012).
\newblock Testing the effect of time pressure on asymmetric dominance and
  compromise decoys in choice.
\newblock {\em Judgment and Decision making}, 7(4):513.

\bibitem[Rendon-Velez et~al., 2016]{rendon2016effects}
Rendon-Velez, E., Van~Leeuwen, P., Happee, R., Horv{\'a}th, I., van~der Vegte,
  W.~F., and De~Winter, J. (2016).
\newblock The effects of time pressure on driver performance and physiological
  activity: a driving simulator study.
\newblock {\em Transportation research part F: traffic psychology and
  behaviour}, 41:150--169.

\bibitem[Roe et~al., 2001]{roe2001multialternative}
Roe, R.~M., Busemeyer, J.~R., and Townsend, J.~T. (2001).
\newblock Multialternative decision field theory: A dynamic connectionist model
  of decision making.
\newblock {\em Psychological Review}, 108(2):370.

\bibitem[Salvia et~al., 2016]{salvia2016effects}
Salvia, E., Petit, C., Champely, S., Chomette, R., Di~Rienzo, F., and Collet,
  C. (2016).
\newblock Effects of age and task load on drivers’ response accuracy and
  reaction time when responding to traffic lights.
\newblock {\em Frontiers in aging neuroscience}, 8:169.

\bibitem[Schwartz et~al., 2002]{schwartz2002maximizing}
Schwartz, B., Ward, A., Monterosso, J., Lyubomirsky, S., White, K., and Lehman,
  D.~R. (2002).
\newblock Maximizing versus satisficing: happiness is a matter of choice.
\newblock {\em Journal of personality and social psychology}, 83(5):1178.

\bibitem[Shi et~al., 2013]{shi2013information}
Shi, S.~W., Wedel, M., and Pieters, F. (2013).
\newblock Information acquisition during online decision making: A model-based
  exploration using eye-tracking data.
\newblock {\em Management Science}, 59(5):1009--1026.

\bibitem[Shimojo et~al., 2003]{shimojo2003gaze}
Shimojo, S., Simion, C., Shimojo, E., and Scheier, C. (2003).
\newblock Gaze bias both reflects and influences preference.
\newblock {\em Nature neuroscience}, 6(12):1317--1322.

\bibitem[Svenson and Maule, 1993]{svenson1993time}
Svenson, O. and Maule, A.~J. (1993).
\newblock {\em Time pressure and stress in human judgment and decision making}.
\newblock Springer Science \& Business Media.

\bibitem[Tavares et~al., 2017]{tavares2017attentional}
Tavares, G., Perona, P., and Rangel, A. (2017).
\newblock The attentional drift diffusion model of simple perceptual
  decision-making.
\newblock {\em Frontiers in neuroscience}, 11:468.

\bibitem[Thomas et~al., 2019]{thomas2019gaze}
Thomas, A.~W., Molter, F., Krajbich, I., Heekeren, H.~R., and Mohr, P.~N.
  (2019).
\newblock Gaze bias differences capture individual choice behaviour.
\newblock {\em Nature human behaviour}, 3(6):625--635.

\bibitem[Uggeldahl et~al., 2016]{uggeldahl2016choice}
Uggeldahl, K., Jacobsen, C., Lundhede, T.~H., and Olsen, S.~B. (2016).
\newblock Choice certainty in discrete choice experiments: Will eye tracking
  provide useful measures?
\newblock {\em Journal of choice modelling}, 20:35--48.

\bibitem[Usher and McClelland, 2004]{usher2004loss}
Usher, M. and McClelland, J.~L. (2004).
\newblock Loss aversion and inhibition in dynamical models of multialternative
  choice.
\newblock {\em Psychological review}, 111(3):757.

\bibitem[Van~Loo et~al., 2018]{van2018using}
Van~Loo, E.~J., Nayga~Jr, R.~M., Campbell, D., Seo, H.-S., and Verbeke, W.
  (2018).
\newblock Using eye tracking to account for attribute non-attendance in choice
  experiments.
\newblock {\em European Review of Agricultural Economics}, 45(3):333--365.

\bibitem[Vij and Walker, 2016]{vij2016and}
Vij, A. and Walker, J.~L. (2016).
\newblock How, when and why integrated choice and latent variable models are
  latently useful.
\newblock {\em Transportation Research Part B: Methodological}, 90:192--217.

\bibitem[Xu et~al., 2018]{xu2018real}
Xu, J., Min, J., and Hu, J. (2018).
\newblock Real-time eye tracking for the assessment of driver fatigue.
\newblock {\em Healthcare technology letters}, 5(2):54--58.

\end{thebibliography}
        
        \end{flushleft}

\end{document}